\documentclass[twocolumn,aps,prl,superscriptaddress]{revtex4-1}
\usepackage{graphicx,subfigure}
\usepackage{amsmath,amssymb,bm}
\usepackage{mathtools}
\usepackage{multirow}
\usepackage{makecell}
\usepackage{textcomp}
\usepackage{float}
\usepackage{xcolor, soul}
\usepackage{color}
\usepackage[normalem]{ulem}
\usepackage{dsfont}
\usepackage{mathrsfs}
\usepackage{braket}
\usepackage{transparent}
\usepackage{xcolor}
\usepackage{hhline}
\usepackage{graphicx}
\usepackage{pdfpages}
\usepackage{bibunits}
\usepackage{filecontents,hyperref}
\usepackage{comment}

\makeatletter
\AtBeginDocument{\let\LS@rot\@undefined}
\makeatother

\graphicspath{ {figures/} }

\makeatletter
\let\saved@includegraphics\includegraphics
\AtBeginDocument{\let\includegraphics\saved@includegraphics}
\makeatother

\makeatletter
\newcommand*{\centerfloat}{%
  \parindent \z@
  \leftskip \z@ \@plus 1fil \@minus \textwidth
  \rightskip\leftskip
  \parfillskip \z@skip}
\makeatother

\usepackage{lineno} 
\begin{document}

\title{Improving the Energy and Angular Resolutions of X-ray Telescopes with Nitrogen-Vacancy Centers in Diamond}

\author{
Ephraim~Gau,$^{1,2,\dag}$
Zhongyuan~Liu,$^{1}$
Henric~Krawczynski,$^{1,2,3}$
Chong~Zu$^{1,3,\ddag}$ 
\\
\normalsize{$^{1}$Department of Physics, Washington University, St. Louis, MO 63130, USA}\\
\normalsize{$^{2}$McDonnell Center for the Space Sciences, Washington University, St. Louis, MO 63130, USA}\\
\normalsize{$^{3}$Center for Quantum Leaps, Washington University, St. Louis, MO 63130, USA}\\
\vspace{0.3cm}
\normalsize{$^\dag$To whom correspondence should be addressed: ephraimgau@wustl.edu}\\
\normalsize{$^\ddag$To whom correspondence should be addressed: zu@wustl.edu}
}

\begin{abstract}
We introduce a focal-plane detector for advancing the energy and angular resolutions of current X-ray telescopes. 
The architecture integrates a metallic magnetic microcalorimeter (MMC) array of paramagnetic absorber pads with a thin layer of nitrogen-vacancy (NV) centers in diamond for simultaneous optical readout. 
An impinging X-ray photon induces a temperature transient in an absorber pad, kept at $\sim$35~mK.
This time- and temperature-dependent magnetic field transient is then optically imaged by diamond NV centers, kept at 4~K and positioned directly below the pad. 
For a 10~$\mu$m absorber length used with a 12~m focal length telescope, our design yields an optimal angular resolution of $\sim$0.17~arcseconds and energy resolution of $\sim$0.70~eV.
Our NV-MMC design improves upon current transition-edge sensors (TES) or MMCs read-out by superconducting quantum interference devices (SQUID) by enabling simultaneous optical readout of the entire MMC array.
Because no additional cryogenic multiplexing electronics are required, our approach scales naturally to larger and finer arrays, supporting finer angular resolutions and wider fields of view.
\end{abstract}

\date{\today}

\maketitle

%%%%%%%%%%%%%%%%%%%%%%%%%%%%%%%%%%%%%%%%%%%%%%%%%%%%%%%%%%%%%%%%%%%%%%%%%%%%%%%%%

\section{Introduction}
\label{s:intro}

The ability to detect X-rays with excellent spatial/angular and energy resolutions is vital for 
properly characterizing astrophysical objects emitting in the X-ray regime (black holes, whether stellar-mass or supermassive, neutron stars, and supernova remnants), 
for aiding in multi-messenger efforts (such as the localization of gravitational wave event counterparts), 
and for fundamental physics endeavors, such as placing constraints on theories of gravity beyond general relativity 
\citep{2013heai.book.....C, 2022hxga.book.....B, 2017ApJ...848L..12A, 2022hxga.book..104S, 2022arXiv221005322B, 2018GReGr..50..100K}. 
The development of new X-ray detectors and missions over the past few decades has enabled increasingly precise detection in both types of resolutions. 
The `High Resolution Camera' (HRC) onboard the {\it Chandra X-ray Observatory} obtains one of the best angular resolutions to date, 0.4~arcseconds, in combination with the 95~eV Full Width at Half Maximum (FWHM) energy resolution at 1.5~keV of another {\it Chandra} instrument, the `Advanced CCD Imaging Spectrometer' (ACIS) detector \citep{1999astro.ph.12097W, 2000SPIE.4012....2W, ChandraCalWebsiteAcis, ChandraCalWebsiteHrc, 2002PASP..114....1W, 2022hxga.book...84W}.
On the other hand, the recently-launched {\it X-Ray Imaging and Spectroscopy Mission} ({\it XRISM}) has a superior energy resolution of 7~eV at 6~keV from its `Resolve' microcalorimeter spectrometer, in combination with an angular resolution of 1.7~arcminutes from its `Xtend' CCD imager. X-rays impinging on the cryogenically cooled absorbers of `Resolve' cause temperature transients in those absorbers, the effect of which is read out by semiconductor thermistors with temperature-dependent resistance \citep{2020arXiv200304962X, 2022arXiv220205399X, 2022SPIE12181E..1TM, 2022SPIE12181E..5ZM, 2023arXiv230301642S}. 
Each state-of-the-art mission thus demonstrates a particular strength in either energy or angular resolution. 
A readout method for X-ray detectors that attains improvements in both directions, then, would be highly desirable. 

As one possible avenue of improvement, state-of-the-art detectors employing microcalorimeters thermally coupled to Transition Edge Sensors (TES) achieve energy resolutions of down to 0.7~eV at 1.5~keV \citep{2005cpd..book...63I, 2005ApPhL..87s4103U, 2009ITAS...19..451S, 2009ITNS...56.2299B, 2012RScI...83i3113B, 2015ApPhL.107v3503L, 2020JLTP..199..949S}.
Complex electronic multiplexing schemes have been developed to read out large numbers of TESs \citep{1999ApPhL..74.4043C, 2001ApPhL..78..371Y, 2010SuScT..23c4004I, 2004ApPhL..85.2107I, 2008ApPhL..92b3514M, 2011PhDT........55M, 2012JLTP..167..707M, 2016ApPhL.109k2604M, 2017ApPhL.111f2601M, 2018JLTP..193..485G}, 
including for forthcoming future X-ray astrophysics missions \citep{2019ITAS...2904472D, 2019JATIS...5b1007B, 2019JATIS...5b1008S, 2020JLTP..199..330S, 2023JATIS...9d1006W}.
Sub-arcsecond imaging over a competitive field of view demands a large number of pixels, and has therefore relied on extensive electronics and multiplexing to date.
For instance, for a focal length of 12~m \citep{2021APh...12602529A}, a 0.5~arcsec angular resolution corresponds to a pixel width of $\sim$30~$\mu$m in the focal plane. 
Thus, obtaining an overall field-of-view similar to the $\sim$30$\times$30 square arcminutes of the {\it Chandra} HRC \citep{2002PASP..114....1W} while keeping such a precise resolution would require the ability to read out a number of pixels on the order of millions.

A closely related detection method is based on (metallic) magnetic (micro)calorimeters (MMCs) with superconducting quantum interference device (SQUID) readouts, offering energy resolutions as good as $\sim1$~eV \citep{1993JLTP...93..709B, 1999PhyB..263..604A, 2000JLTP..118....7F, 2000JLTP..121..137E, 2000NIMPA.444..100F, 2000NIMPA.444..211S, 2000SPIE.4140..445A, 2003PhyB..329.1594F, 2003RScI...74.3947F, 2003SuScT..16.1404Z, 2004NIMPA.520...27F, 2004NIMPA.520...52Z, 2004PSSCR...1.2824F, 2004SPIE.5501..346B, 2005cpd..book..151F, 2005ITAS...15..773S, 2006JAP....99hB303Z, 2008JLTP..151..345S, 2008JLTP..151..357H, 2008JLTP..151..363S, 2008JLTP..151..337B, 2008SPIE.7021E..1KB, 2009AIPC.1185..571F, 2009AIPC.1185..579B, 2009AIPC.1185..583R, 2009ITAS...19...63F, 2009JPhCS.150a2013F, 2012JLTP..167..254B, 2012JLTP..167..269P, 2014JLTP..176..617P, 2016JLTP..184..108L, 2016JLTP..184..344K, 2017AIPA....7a5007K, 2017SuScT..30f5002K, 2018JLTP..193..365K, 2018JLTP..193..462W, 2018JLTP..193..668S, 2019ITAS...2908143Y, 2019JATIS...5b1009S, 2020JLTP..199..916Y, 2021JInst..16P8003M, 2022JLTP..209..337D, 2023ITAS...3359334B, 2024ApPhL.124c2601K}. 
However, these devices also rely on elaborate cryogenic/electronic multiplexing to read-out large pixel arrays.

A scheme that minimizes cryogenic electronics \footnote{And, thus, minimizing complications involved in fabricating and manipulating the wires and SQUIDs necessary for the desired sizes and numbers of absorbers. Additionally, SQUIDs may also only tolerate magnetic fields of certain strengths, and are also more sensitive to unintended inductances and currents.} and reduces thermal links between cryogenic and room temperature stages will be ideal for scaling to larger numbers of absorbers and hence finer angular resolutions and/or a wider fields of view.
To address these constraints, we propose an alternative MMC readout that uses nitrogen-vacancy (NV) centers in diamond.

In recent years, NV centers have emerged as precision nanoscale quantum sensors of magnetic and electric fields, strain, and temperature across diverse environments,
with applications ranging from material characterization and biological imaging to the search for dark matter candidates and new particles \citep{2014ARPC...65...83S, doherty2013nitrogen, 2019Nanop...8..209L, 2013Natur.500...54K, 2017NanoL..17.1496B, 2017Natur.549..252G, 2018PhRvL.121x6402M, 2019ApPhL.114w1103W, hsieh2019imaging, block2021optically, su2021search, 2024Natur.627...73B, jiao2021experimental, glenn2017micrometer, thiel2019probing, ariyaratne2018nanoscale, ajoy2015atomic, rajendran2017method, chigusa2023light, kashem2025multiplexed}.
In our architecture, the NV-based optical readout is projected to deliver energy and angular resolutions competitive with those of current TES or MMC arrays, while simultaneously offering a readily scaled-up pixel array for further improving the angular resolution and field-of-view.
Additionally, the imaging-based optical readout allows simultaneous sampling of all pixels, reducing ambiguities from crosstalk that may occur in other setups.

\begin{figure*}[tb]
\begin{center}
    \includegraphics[width=.95\linewidth]{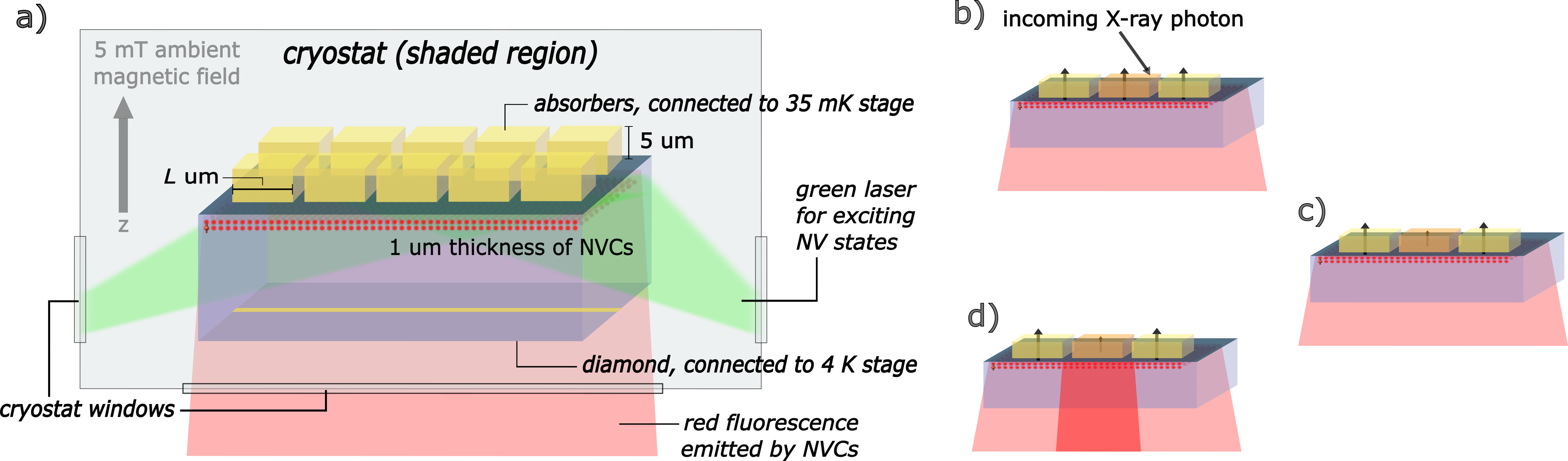}
\end{center}
\vspace*{-0.3cm}
\caption{
{\bf Schematic of the NV-MMC setup, and of its detection procedure.}
{\bf (a)} An array of erbium-doped gold absorbers, each of dimension $L~\mu$m $\times$ $L~\mu$m $\times$ 5~$\mu$m, is supported by thin layers of graphite and silver 100~nm above the diamond plate.
The diamond contains a dense ensemble of NV sensors (red dots) within the top 1~$\mu$m of the diamond, produced via an ion implantation and annealing process.
The paramagnetic absorber pads are cooled to $\sim$35~mK, whereas the diamond stage is connected to the $\sim$4~K stage; the overall cryostat is indicated by the shaded box. 
The green excitation laser (see Fig.~\ref{fig:fig3}) for the NV centers is delivered in from one side window of the cryostat, undergoes total internal reflection at the top surface of the diamond, and leaves the cryostat through another window on the opposite side.
The NV fluorescence then exits the cryostat via another window at the bottom of the cryostat, where it is focused by an objective lens onto an imaging CCD camera below the lens.
The placement of the CCD outside the cryostat greatly reduces the heat load that would otherwise need to be cooled by the cryogenics. 
For optimal absorber performance, the interior of the cryostat maintains a 5~mT field.
The gold strip for delivering the needed microwave power (also see Fig.~\ref{fig:fig3}) is located at the bottom of the diamond.
{\bf (b)-(d)} A zoomed-in schematic, showing each step of the detection process. 
An incident X-ray photon strikes and heats an absorber pad. 
The paramagnetic absorber, having a temperature-dependent magnetization, thus produces a magnetic transient. 
This magnetic field change shifts the NV center energy transition frequencies and thus alters the fluorescence produced by the NV centers below the relevant pad (Fig.~\ref{fig:fig2}b and c). 
By imaging that fluorescence as a function of time and frequency with a CCD camera, the incident X-ray event can thus be sensed with high spatial and energy resolution.
}
 \label{fig:fig1}
\end{figure*}

Our setup comprises an array of micrometer-scale erbium-doped gold absorber-microcalorimeter pads positioned $\sim$100~nm above a diamond plate, with a thin layer of NV centers embedded within the top $\sim$1~$\mu$m of the diamond (Fig.~\ref{fig:fig1}). 
When an incident X-ray strikes an absorber pad, the absorbed photon raises the pad temperature and thus changes its magnetization. 
The corresponding change in magnetic field is then sensed \emph{in-situ} via the change in the optically-detected electron-spin resonance (ODMR) spectrum of the NV centers. 

The rest of the paper is organized in the following manner.
We first describe the detector architecture and principles.
We then compute the projected energy and angular resolutions of the NV-MMC. 
We conclude with potential implementations and improvements to reach the optimal performance of the NV-MMC.
Additional derivations and mitigation strategies---particularly for optical and thermal loads---are provided in the Supplementary Material.

%%%%%%%%%%%%%%%%%%%%%%%%%%%%%%%%%%%%%%%%%%%%%%%%%%%%%%%%%%%%%%%%%%%%%%%%%%%%%%%%%

\section{Proposed experimental setup}
\label{s:det}

Our concept builds on MMC detectors
\citep{1993JLTP...93..709B, 1999PhyB..263..604A, 2000JLTP..118....7F, 2000JLTP..121..137E, 2000NIMPA.444..100F, 2000NIMPA.444..211S, 2000SPIE.4140..445A, 2003PhyB..329.1594F, 2003RScI...74.3947F, 2003SuScT..16.1404Z, 2004NIMPA.520...27F, 2004NIMPA.520...52Z, 2004PSSCR...1.2824F, 2004SPIE.5501..346B, 2005cpd..book..151F, 2005ITAS...15..773S, 2006JAP....99hB303Z, 2008JLTP..151..345S, 2008JLTP..151..357H, 2008JLTP..151..363S, 2008JLTP..151..337B, 2008SPIE.7021E..1KB, 2009AIPC.1185..571F, 2009AIPC.1185..579B, 2009AIPC.1185..583R, 2009ITAS...19...63F, 2009JPhCS.150a2013F, 2012JLTP..167..254B, 2012JLTP..167..269P, 2014JLTP..176..617P, 2016JLTP..184..108L, 2016JLTP..184..344K, 2017AIPA....7a5007K, 2017SuScT..30f5002K, 2018JLTP..193..365K, 2018JLTP..193..462W, 2018JLTP..193..668S, 2019ITAS...2908143Y, 2019JATIS...5b1009S, 2021JInst..16P8003M, 2022JLTP..209..337D, 2023ITAS...3359334B, 2024ApPhL.124c2601K},
but replaces the SQUID electronics with optically addressable NV centers to read out the demagnetization transients produced when X-ray absorption heats the paramagnetic pads, in addition to using a combined absorber-thermometer.

Figure~\ref{fig:fig1} displays the overall detector architecture. 
An array of erbium-doped gold (incorporating enriched $^{166}$Er at a concentration of 300~ppm in gold \cite{2005cpd..book..151F}) absorbers faces the incoming X-rays, with each pad of dimension $L ~ \times ~ L ~ \times ~ 5~\mu$m, where $L$ ranges from 1 to 20~$\mu$m. 
The concentration of erbium ($\sim$300~ppm) is chosen so as to not overly increase the spin-spin Ruderman-Kittel-Kasuya-Yosida (RKKY) exchange interaction \citep{1993JLTP...93..709B} (for more on the choice of material, also see \cite{1999PhyB..263..604A, 2000JLTP..118....7F, 2000JLTP..121..137E, 2003RScI...74.3947F}).
The thickness of 5~$\mu$m is chosen to surpass the characteristic absorption length in gold (4.4~$\mu$m) for a 10~keV X-ray \citep{NIST}. 
These gold-erbium absorbers can be produced in sputtering methods similar to those described in \cite{2004NIMPA.520...52Z, 2005cpd..book..151F, 2006JAP....99hB303Z, 2008JLTP..151..357H, 2008SPIE.7021E..1KB, 2009AIPC.1185..571F}, or with vapor deposition-type techniques as in \cite{2000SPIE.4140..445A, 2004NIMPA.520...52Z, 2005cpd..book..151F, 2006JAP....99hB303Z}, with each of these sources producing detectors of a similar magnitude of thickness to (or even thinner than) what we require.

These absorber pads will be kept at 35~mK by the mixing chamber stage of a dilution refrigerator, as well as at ambient magnetic fields of $\sim$5.1~mT, to take advantage of the optimized combination of the heat capacity of gold and the magnetization of the erbium-doped gold at those conditions \citep{2005cpd..book..151F}.
Erbium doping makes the alloy paramagnetic, with a magnetization $M$ approximately linearly dependent on temperature $T$,
\begin{equation}
M \simeq - \alpha T,
\end{equation} where $\alpha>0$ is the constant of proportionality. 
Absorption of an X-ray photon raises the pad temperature by $\delta T$, reducing its magnetization and hence the local magnetic field generated by that pad \citep{2006JAP....99hB303Z}. 
By measuring the field change, the energy of the incident X-ray can be determined.

To detect this magnetic field change \emph{in situ}, we use an NV-based wide-field magnetometer comprising a [111]-cut diamond plate ($1~\mathrm{mm}\times1~\mathrm{mm}\times0.1~\mathrm{mm}$) mounted $\sim$100~nm beneath the absorbers. 
This diamond, kept at the 4~K stage, contains a thin layer of NV centers in its top $\sim$1~$\mu$m.
To optically excite the NV centers, a 532~nm green laser is applied to the diamond through a side window into the cryostat.
Due to the large refraction index of diamond ($n\sim2.4$), the laser beam can be aligned to undergo total internal reflection at the top diamond surface and then exit the cryostat through the opposite side window, thereby minimizing laser heating of cryogenic stages (Fig.~\ref{fig:fig1}a) \citep{2019Nanop...8..209L}.
The down-traveling red fluorescence from the NV centers is collected by an objective lens outside the cryostat and imaged onto a CCD camera (Fig.~\ref{fig:fig1}b).

In terms of the region between the absorbers and the diamond:
the absorbers will be deposited atop a layer of graphite (AGOT), of thickness $\sim$50~nm, chosen for its low thermal conductivity \citep{2007mmlt.book.....P}.
This layer serves as the thermal separation between the 35~mK and 4~K stages.
Below that, a layer of silver, of thickness $\sim$50~nm, will be deposited atop the diamond. 
This silver layer reflects away the up-traveling red NV fluorescence from hitting the paramagnetic absorber pads \citep{1962PhRv..128.1622E, 1961PhRv..121.1100T}.

\subsection{NV centers as magnetic field sensors}

\begin{figure*}[tb]
\begin{center}
    \includegraphics[width=.95\linewidth]{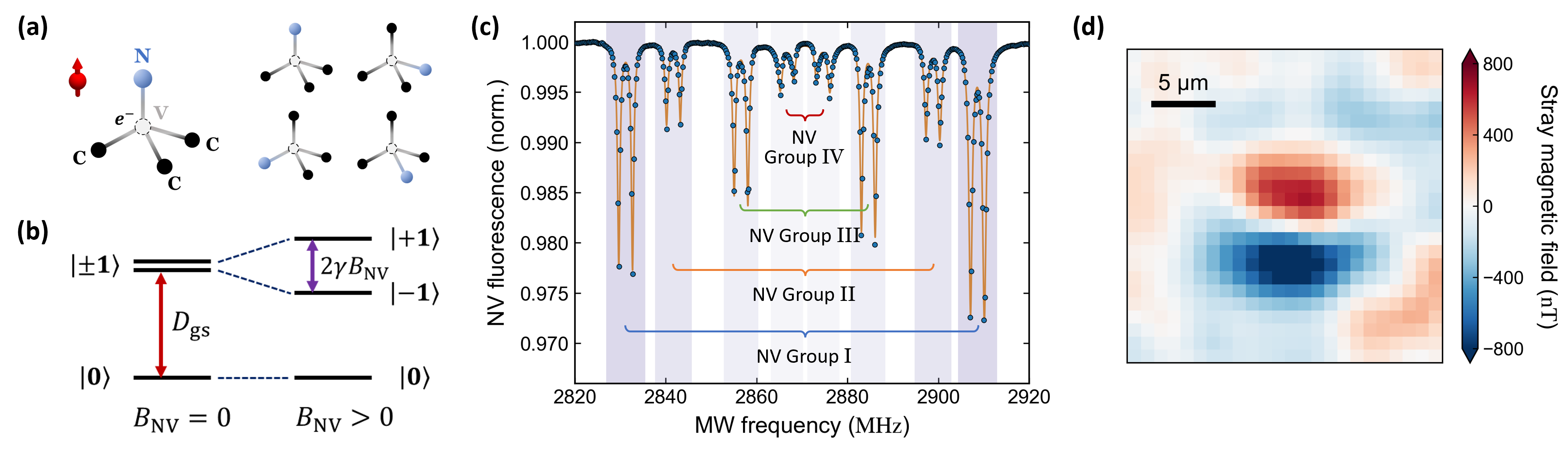}
\end{center}
\vspace*{-0.3cm}
\caption{{\bf Principle, and demonstration, of magnetic field sensing with NV centers.}
{\bf (a)} The lattice structure of NV centers in diamond: two adjacent carbon atoms (black) in the lattice are replaced by a nitrogen atom (light blue) and a vacancy (dotted circle). 
In a single-crystal diamond, there are thus four possible orientations of the NV crystallographic axis.
{\bf (b)} Energy levels of (spin-1) NV centers: in the absence of a magnetic field, the system exhibits only a zero-field splitting of $D_\mathrm{gs} = 2870$~MHz between the $|m_s = 0\rangle$ and $|m_s = \pm1\rangle$ sublevels.
By applying an external magnetic field, the degeneracy between the $|m_s = \pm1\rangle$ sublevels is lifted due to the Zeeman effect, leading to an additional splitting of $2\gamma B_\mathrm{NV}$, where $B_\mathrm{NV}$ is the projection of the external magnetic field onto the nitrogen-vacancy (NV) bond axis.
{\bf (c)} An example of the electron spin resonance (ESR) spectrum of a NV ensemble with an external magnetic field of approximately 3~mT.
The spectrum is divided into four groups of resonances (marked by the shaded areas), with each group corresponding to one of the four NV crystallographic axes. 
The splitting exhibited by each group is determined by the magnitude of the projection of the external magnetic field onto the respective NV axis. 
(We note that the resonance peaks within each shaded area also exhibit a small hyperfine splitting of about 3~MHz, originating from the hyperfine interaction between the NV center and the $^{15}$N nuclear spin~\citep{smeltzer2009robust}.)
{\bf (d)} A wide-field magnetic field image of the dipole due to a magnetic bead deposited on top of a diamond surface, acquired with our NV diamond microscope. 
The image demonstrates sub-$\mu$T precision and has a spatial resolution of $\sim$~1~$\mu$m after applying a Gaussian filter with $\delta=1.5~\mu$m to reduce background noise.
}
 \label{fig:fig2}
\end{figure*}

An NV center consists of a substitutional nitrogen atom and an adjacent vacancy in a diamond lattice (see Fig.~\ref{fig:fig2}a, which also displays all four possible crystallographic orientations of the NV centers in a single-crystalline diamond).
The electronic spin energy levels of the NV center exhibit a spin-1 triplet state ($|m_s = 0,\pm1\rangle$), whose three sublevels are nearly equally populated at thermal equilibrium (Fig.~\ref{fig:fig2}b). 
The NV center can be initialized to the $|m_s = 0\rangle$ state via an optical pumping process through excitation by the green laser.
This process also results in a direct probe of the spin state: the $|m_s=0\rangle$ sublevel, in relaxing from excitation by the laser, emits more red fluorescent photons than the $|m_s=\pm1\rangle$ sublevels.
When a microwave field resonant with the $|m_s=0\rangle \leftrightarrow |m_s=\pm1\rangle$ transition is applied, some of the spin population is transferred from the bright states ($|m_s=0\rangle$) to the dim states ($|m_s=\pm1\rangle$), causing a measurable drop in fluorescence.
As a result, by sweeping the applied microwave field through a range of frequencies and monitoring the NV fluorescence on a CCD camera, one can determine the characteristic resonant frequencies of the NV centers.
This is known as optically detected magnetic resonance (ODMR) spectroscopy~\citep{2014ARPC...65...83S, 2019Nanop...8..209L, doherty2013nitrogen}.

Then, the sensitivity of the ODMR spectrum to external magnetic fields permits NV centers to be used as a magnetometer.
In the absence of magnetic fields, the $|m_s=\pm1\rangle$ states are degenerate and separated from the $|m_s = 0\rangle$ state by merely a zero-field splitting of $D_\mathrm{gs} = 2.87$~GHz, resulting in a single-resonance ODMR spectrum.
However, when an external magnetic field is present, the degeneracy between the $|m_s=\pm1\rangle$ sublevels is lifted via the Zeeman effect, leading to a splitting of $\Delta = 2\frac{\gamma}{(2\pi)} B_\mathrm{NV}$ between the $|m_s=\pm1\rangle$ sublevels, where $\gamma = (2\pi) \times 2.8025$~MHz/G is the electron gyromagnetic ratio and $B_\mathrm{NV}$ is the magnetic field projection along the nitrogen-vacancy bond axis. 
Thus, a typical ODMR spectrum of NV centers from a single crystallographic group exhibits two distinct resonance dips. 
The separation between these resonances---increasing when the external magnetic field increases, and decreasing when it decreases---can then be used to precisely determine the changes in the strength of the environmental magnetic field \citep{2014ARPC...65...83S, doherty2013nitrogen, 2019Nanop...8..209L}.

Figure~\ref{fig:fig2}c shows an example of the ODMR spectrum measured using the diamond sample proposed in this work.
A few remarks are in order.
First, since a dense ensemble of NV defects is used, within an optical detection volume ($\sim$1~$\mu$m$^3$), there exist all four (Fig.~\ref{fig:fig2}a) different crystallographic groups of NV centers.
As a result, the corresponding ODMR spectrum displays four pairs of resonances.
Second, there is a small splitting of about 3.1~MHz within each of the eight resonance dips, arising from the hyperfine coupling between the NV center and proximal $^{15}$N nuclear spin~\citep{smeltzer2009robust}.
Third, the spatial resolution of our NV magnetometer is optically diffraction-limited (around 1~$\mu$m).
Figure~\ref{fig:fig2}d shows a measured magnetic field map of a single magnetic nanobead placed on top of the diamond surface.
Using an 10$\times$ objective lens, a magnetic field map with spatial resolution of $\sim$1~$\mu$m and magnetic field precision of $\sim$100~nT is achieved here.

%%%%%%%%%%%%%%%%%%%%%%%%%%%%%%%%%%%%%%%%%%%%%%%%%%%%%%%%%%%%%%%%%%%%%%%%%%%%%%%%%

\section{Projected performance of the NV-MMC}
\label{s:res}

\subsection{Energy resolution}

\subsubsection{Temperature transients caused by an incident X-ray}

To estimate the energy resolution of the NV-MMC detector, we first characterize the temperature increase, $\delta T$, of an erbium-gold absorber pad when an incident X-ray photon with energy $\delta E$ arrives.
The heat capacity of the erbium-doped gold pad at temperature $\sim$35~mK and ambient magnetic field of $\sim$5~mT has been measured to be $\sim$$3.6 \times 10^{-4} ~{\rm J}~{\rm K}^{-1}~{\rm mol}^{-1}$ (from Fig.\,2 of \cite{2005cpd..book..151F}).
Consequently, taking an absorber pad width of $L = 10~\mu$m as an example, the temperature increase per unit X-ray energy is given by $dT/dE \approx 9.3 \, \mathrm{mK/keV}$.

The graphite plate below the absorbers will cool each pad after an X-ray hits. 
If similar strengths of thermal couplings/Kapitza resistances as in \cite{2005cpd..book..151F, 2009AIPC.1185..571F, 2009ITAS...19...63F} are used, it is possible for the absorber to take $\sim$ 1 to 10~ms (or, indeed, even longer) to return to the base temperature (35~mK). 
Thus, in the following analysis, the corresponding temperature increase due to an incident X-ray is taken to be a constant pulse lasting for $\delta t = 1$~ms.

\subsubsection{Magnetic field transients caused by a given change in temperature}

We next examine the resulting magnetization change of the absorber, $\delta M$, given a temperature increase of $\delta T$.
The magnetization as a function of temperature in the erbium-doped gold is influenced by many factors, including the indirect RKKY interaction, as well as the nuclear spin of the erbium ion \citep{2005cpd..book..151F}.
Based on the experimental measurements and numerical simulations from \citep{2005cpd..book..151F}, the slope that one obtains for the magnetization versus temperature curve is about $\partial M/\partial T = -$4.1~A/m~mK$^{-1}$.

The resulting magnetic field change at different locations of the NV sensor layer, $\delta B$, can be calculated using
\begin{equation}
\delta \bm{B}(\bm{r}) = \frac{\mu_0}{4\pi}  \int \frac{3\hat{\bm{r}}~(\delta \bm{M}\cdot \hat{\bm{r}}) - \delta \bm{M}}{r^3} dV,    
\end{equation}
where $\mu_0$ is the vacuum permeability, and $\bm{r}= r ~ \hat{\bm{r}}$ (with $|\hat{\bm{r}}|=1$) is the vector pointing from the differential absorber volume element towards the NV centers as the whole volume of a given pad is integrated over. 

\begin{figure}[tb]
\begin{center}
    \includegraphics[width=0.95\linewidth]{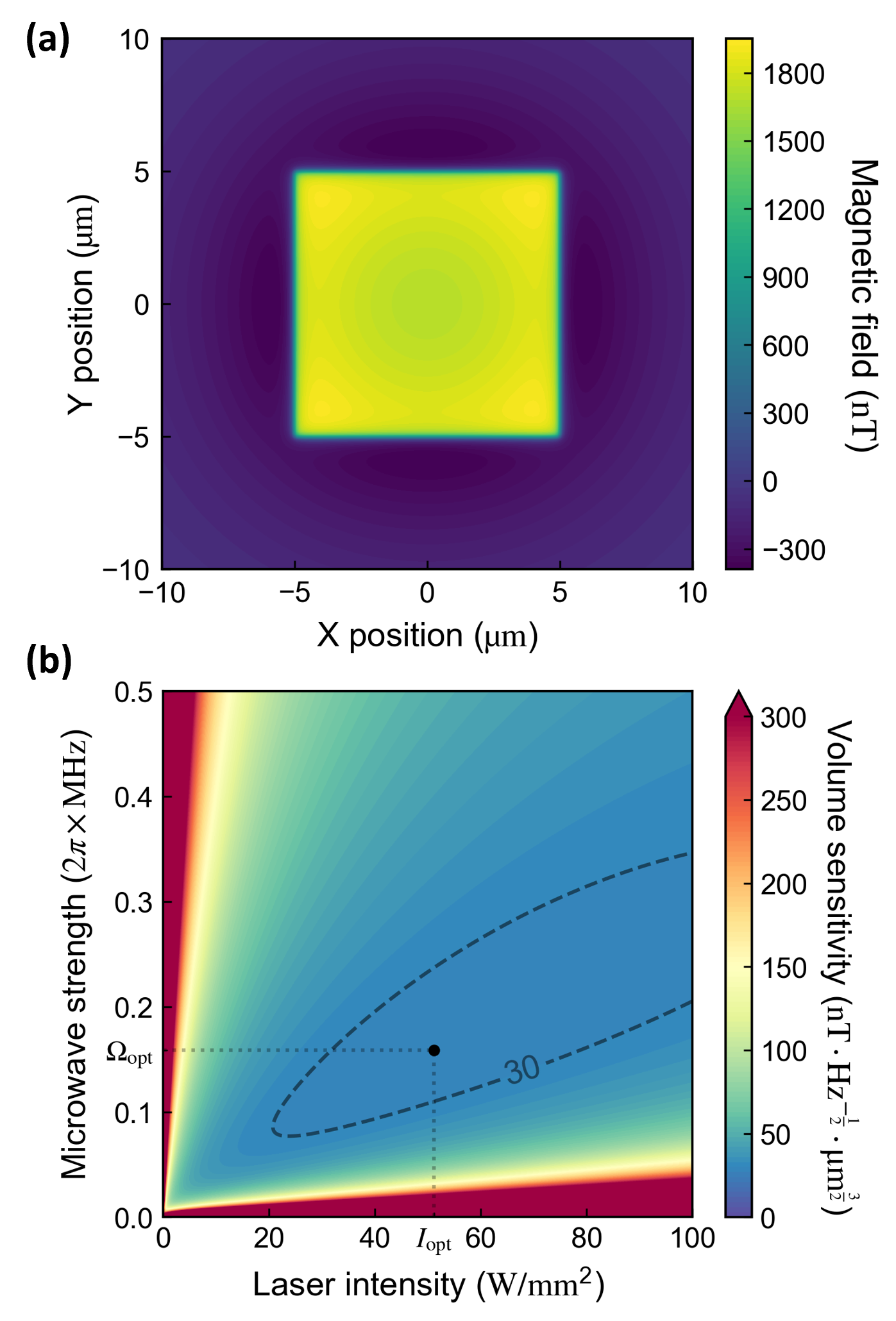}
\end{center}
\vspace*{-0.3cm}
\caption{
{\bf Magnetic field sensitivity calculations for the NV-MMC.}
{\bf (a)} The calculated distribution of magnetic field change in the $z$-direction at the location of the NV centers, $100$~nm beneath a $10~\mu \rm m \times 10~\mu \rm m \times 5~\mu \rm m$ absorber pad, given a $1$~mK increase in the  temperature of that absorber.
{\bf (b)} The calculated magnetic field volume sensitivity of the NV ensemble, using the sample parameters listed in Table~\ref{tab:SampleParameters}, as a function of microwave (MW) strength and laser intensity. 
The optimal sensitivity, 27~${\rm nT}~{\rm Hz}^{-\frac{1}{2}}~\mu$m$^{\frac{3}{2}}$, is marked with the black dot at $I_{\mathrm{opt}}\approx 51~\mathrm{W/mm^2}$ and $\Omega_{\mathrm{opt}}\approx (2\pi)\times0.16~$MHz. 
The black dashed circle displays the contour for all parameter combinations that yield a sensitivity of 30~${\rm nT}~{\rm Hz}^{-\frac{1}{2}}~\mu$m$^{\frac{3}{2}}$ or better. 
Thus, the laser intensity and microwave strengths can both be roughly halved without significantly changing the sensitivity. 
}
 \label{fig:fig3}
\end{figure}

Combining the above two steps, Fig.~\ref{fig:fig3}a displays, for a $\delta T = 1$~mK, the spatial distribution of the corresponding $\delta B$ (projected along the vertical or z-direction shown in Fig.~\ref{fig:fig1};
this direction is also the direction of a nitrogen-vacancy bond axis in our $[111]$-cut diamond).
For each given pad dimension $L$, a relevant detection volume of NV centers beneath ($V_\mathrm{r}$) is determined (see Supplementary Materials for more details). 
The average change in magnetic field across the detection volume yields, for example, $dB/dT \approx 1500 \, \mathrm{nT/mK}$ for $L = 10~\mu$m.

\subsubsection{Magnetic field sensitivity of NV centers}

We now turn to estimating the magnetic sensitivity, $\delta B_{min}$ (the smallest magnetic difference that could be sensed by the NV centers within a given time $\delta t$). 
The shot-noise-limited magnetic field sensitivity for the NV ODMR measurement is given by \cite{2019Nanop...8..209L}
\begin{align}
    \eta \approx \frac{8\pi}{3\sqrt{3}} \frac{1}{\gamma} \frac{\delta\nu}{C \sqrt{R}} ,  \label{eq:sensitivity}
\end{align}
where $\delta\nu$ is the FWHM of ODMR resonance dip (assuming a Lorentzian functional form), $C$ is the normalized contrast of the resonance~\footnote{The contrast $C$ refers to the percent change in fluorescence between the ambient value of fluorescence and the decreased fluorescence at the bottom of the Lorentzian dip of the resonance. For example, the left peaks in Group I of Fig.~\ref{fig:fig2}c show contrasts of a little less than 2.5~\%.}, $R$ is the photon count rate, and $\gamma$ is the same electron gyromagnetic ratio as before.
$\delta\nu$, $C$, and $R$ are highly dependent on the substitutional nitrogen ([N]) and nitrogen-vacancy defect ([NV]) concentrations in the diamond, as well as on the laser intensity $I$ and the microwave field driving strength $\Omega$.
Considering a diamond sample achievable by current state-of-the-art fabrication methods (Table~\ref{tab:SampleParameters}), we calculate the volume sensitivity as a function of $I$ and $\Omega$ \citep{dreau2011avoiding}, to yield Fig.~\ref{fig:fig3}c (detailed calculations are described in the Supplementary Materials).

The optimal magnetic field sensitivity of $\eta_{\rm opt} \approx 27~{\rm nT}~{\rm Hz}^{-\frac{1}{2}}~\mu$m$^{\frac{3}{2}}$ is achieved with a laser intensity of $I \approx 51~{\rm W/mm^2}$ and microwave driving strength of $\Omega \approx (2\pi) \times 0.16$~MHz. 
Accounting for the aforementioned duration of the temperature transient from an X-ray event ($\delta t=1~$ms) and the relevant detection volume of NV centers underneath a given pad, we then arrive at the minimum detectable magnetic field of our NV-MMC.
For instance, at $L=10~\mu$m, the magnetic field sensitivity is estimated to be $\delta B_\mathrm{min} = \eta_\mathrm{opt}/(\sqrt{\delta t} \sqrt{V_\mathrm{r}}) \approx 110$~nT. 

\begin{table}[h!]
%  \centerfloat
    \begin{tabular}{ c|c|c|c|c|c } 
     \hline
      [N] & [NV] & [NV]/[N] & $R_\mathrm{sat}$ & Contrast$_{\mathrm{max}}$ & FWHM \\ \hline
     $20$~ppm & $4.5$~ppm & $22.5\%$ & $10^5$~/~s & $7.5\%$ & $1$~MHz \\
     \hline
    \end{tabular}
  \caption{
  {\bf Diamond sample parameters.} [N] and [NV] characterize the density of N and NV impurities in the proposed diamond sample. 
  $R_\mathrm{sat}$ is the count rate for a single NV center when addressed by the laser at saturation power. 
  Contrast$_{\mathrm{max}}$ is the maximum amplitude of the ESR spectrum dips, relative to the baseline fluorescence level. 
  The full-width-at-half-maximum (FWHM) denotes the linewidth of each individual ESR peak. 
  Particle densities are noted in parts per million (ppm); $1$~ppm corresponds to $1.76\times 10^5~\mathrm{\mu m}^{-3}$ in diamond. 
  Sources: \cite{2019Nanop...8..209L, 2020RvMP...92a5004B, he2023quasi}.
  }
     \label{tab:SampleParameters}
\end{table}

\subsubsection{Overall energy resolution of the NV-MMC}

All in all, the three steps of estimating the energy resolution can be conceptually summarized through the following equation:
\begin{align}
    \delta E_\mathrm{min} = \frac{dE}{dT} \frac{dT}{dB} \delta B_\mathrm{min} \,\,   \label{eq:key}.
\end{align}
Again using the example of a pad length $L=10~\mu$m, we obtain an energy resolution $\delta E_\mathrm{min} \approx 7.5 \, \mathrm{eV}$ (labeled as ``NV-MMC'' in Fig.~\ref{fig:fig4}). 

\subsubsection{Potential optimizations for the energy resolution}

Two methods to further improve the energy resolution of the NV-MMC are now discussed.

We start with improving the photon count rate $R$ in Eqn.~(3), which affects the $\eta_\mathrm{opt}$ and thus the $\delta B_\mathrm{min}$ of Eqn.~(4).
Owing to diamond's high refraction index, a substantial fraction of the down-traveling fluorescent photons reflect at the bottom diamond surface, limiting the photon count rate at the CCD.
To mitigate this, prior work has shown that by fabricating the bottom surface of the diamond into spherical solid immersion lenses, one can significantly enhances the photon collection rate by a factor of 10, thus boosting the energy resolution by a factor of $\sim$$\sqrt{10}$ \citep{2010ApPhL..97x1901H}.

The second strategy is to increase the temperature transient timescale $\delta t$. Given that \cite{2005cpd..book..151F} lists a variety of possible decay times (from fractions of a millisecond to tens of milliseconds), we can envision increasing $\delta t$ by worsening the thermal coupling between the absorbers and the layer of graphite beneath. 
Having $\delta t \approx10$~ms would then improve the energy resolution by yet another factor of $\sim$$\sqrt{10}$.

\begin{figure}[tb]
% \begin{center}
    \includegraphics[width=1.05\linewidth]{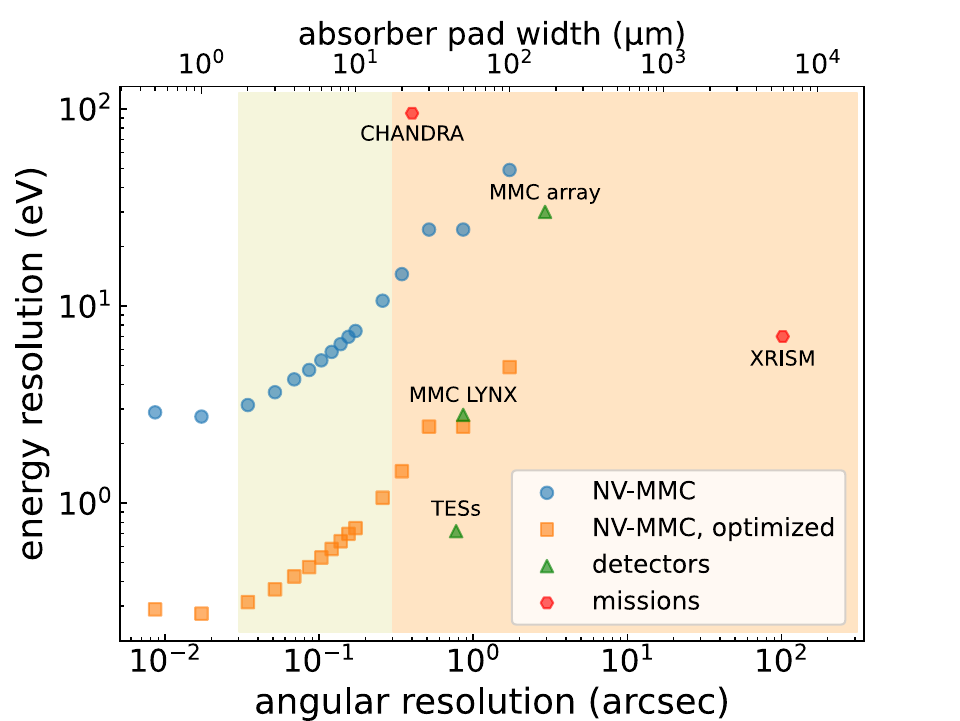}
% \end{center}
\vspace*{-0.3cm}
\caption{{\bf Energy and angular resolutions of the NV-MMC, compared to other missions and detectors.} Energy resolutions as a function of absorber pixel width (top horizontal axis) and angular resolution (bottom horizontal axis) for the NV-MMC, with the absorber pads kept at a distance of 100~nm above the diamond. These results are compared with those from some current state-of-the-art X-ray detectors (data for other missions/detector designs are taken from \cite{ChandraCalWebsiteAcis, 2022arXiv220205399X, 2015ApPhL.107v3503L, 2016JLTP..184..344K, 2020JLTP..199..916Y}). The {\it Chandra} energy resolution is given for 1.5~keV; the {\it XRISM} and MMC energy resolutions are given for 6~keV; and the TES energy resolution is given for 1.5~keV. For test arrays not yet fully implemented on missions, their angular resolution is given assuming a focal length of 12~m. The darker (lighter) shaded region indicates an ability to spatially differentiate an angular resolution on the scale of one-thousandth (one-ten-thousandth) the rough angular extent of the Crab Nebula, a pulsar wind nebula and prototypical high-flux source for X-ray astronomy.
}
 \label{fig:fig4}
\end{figure}

Combining these two improvements would allow an optimized energy resolution of $\delta E_\mathrm{opt} \approx 0.75$~eV for $L = 10~\mu$m (labeled as ``NV-MMC, optimized'' in Fig.~\ref{fig:fig4}), comparable to current state-of-the-art X-ray detectors.

Considering that absorber of different sizes will have differing energy resolutions, we also investigate the results as a function of the absorber pad length $L$. 
Fig.~\ref{fig:fig4} plots the energy resolution of the NV-MMC as a function of absorber width (which, for our purposes, correlates directly with angular resolution; see the relevant section below). 
Two main effects are at play. 
The first term in Eqn.~\ref{eq:key} (dE/dT, the heat capacity of the absorber pad) grows quadratically with $L$ ($\sim L^2\times5~\mu$m, since the pad thickness is fixed).
Simultaneously, the last term in Eqn.~\ref{eq:key} ($\delta B_\mathrm{min}$, the minimum detectable magnetic field) scales with $L^{-3/2}$ due to the change of the NV sensor detection volume ($\delta B_\mathrm{min} \sim 1/\sqrt{V_{r}}$, while $V_\mathrm{r}\sim L^3$).
Thus, the effect of the former dominates, and the energy resolution generally improves with decreasing absorber size $L$ (Fig.~\ref{fig:fig4}) \footnote{The second term in equation \ref{eq:key}, dT/dB, also improves with smaller absorber size (on average, improving with the square root of $L$) until about 2 or 3~$\mu$m, after which it starts worsening with decreasing absorber size. However, we only see this slowly take effect, leading to the turnover being at around 1~$\mu$m, as it needs to override the main trend previously described.}.

%%%%%%%%%%%%%%%%%%%%%%%%%%%%%%%%%%%%%%%%%%%%%%%%%%%%%%%%%%%%%%%%%%%%%%%%%%%%%%%%%

\subsection{Angular resolution}

Compared with conventional SQUID readouts for MMCs and TESs, our NV-MMC platform uses wide-field optical interrogation with minimal electronics. 
This enables much denser arrays with smaller absorber pads, thereby substantially improving angular resolution.

For a single incident X-ray on one absorber (Fig.~\ref{fig:fig3}a), the induced heating and the accompanying magnetic field change remains largely confined, in the NV layer, to the area directly beneath that pad.
Therefore, through taking constant optical images of the area under the entire array and thus monitoring any NV fluorescence change, we can determine when and which particular absorbers are struck by X-rays. 

Then, in placing the NV-MMC behind a telescope with a focal length of $f = 12~$m, similar to that of a recent balloon-borne experiment \citep{2021APh...12602529A}, the resulting angular resolution is estimated (using the small angle approximation) by
\begin{align}
    \delta a = \frac{L}{f}.  \label{eq:focallengthconversion}
\end{align}
Once again taking $L = 10~\mu m$ as an example, we obtain an angular resolution for the NV-MMC of $\sim$0.17~arcseconds (at that focal length), already more precise than existing missions and detector setups (Fig.~\ref{fig:fig4}). 
By decreasing $L$ to $1~\mu$m, the angular resolution can be further improved by another order of magnitude to $\sim$0.017~arcsec.
Practically, the minimum $L$ would be around $1~\mu$m, roughly the optical diffraction limit of the NV readout. % due to wavelength

%%%%%%%%%%%%%%%%%%%%%%%%%%%%%%%%%%%%%%%%%%%%%%%%%%%%%%%%%%%%%%%%%%%%%%%%%%%%%%%%%

\section{Conclusions and outlook}
\label{s:dis}

The optimized NV-MMC X-ray detector has a projected energy resolution as low as $\sim$0.30~eV and angular resolution as precise as $\sim$0.017~arcsec when used with a telescope of focal length 12~m, as competitive as or even potentially surpassing the resolutions of current missions and developed detector setups (Fig.~\ref{fig:fig4}).
Looking forward to next steps, our proposed detector opens the door to several intriguing and readily-accessible paths of development.

First, developing and launching the balloon-borne X-ray missions {\it XL-Calibur} \cite{2021APh...12602529A, 2020ApJ...891...70A, 2022APh...14302749A} and {\it DR-TES} (`Dilution Refrigerator --- Transition Edge Sensor') \footnote{Gau et al., in progress} \citep{2020apra.prop...44K, 2023JATIS...9b4006S} provides a clear pathway to a proof-of-principle flight that integrates the NV-MMC detector and its cryogenic infrastructure.
In parallel, established methods for wide-field NV magnetometry de-risk the optical readout and calibration pipeline.

A suitable next step for the NV-MMC would then be to manufacture the absorber array (deposition machines for microfabrication of the absorbers are readily accessible) and diamond and attempt in-lab tests, before eventually attempting a similar test flight as {\it DR-TES}. 
For such a setup, the CCD camera, laser, and X-ray test source would be positioned just outside a cryostat with windows, and the NV-MMC absorber array and diamond, both cooled by a mini-DR from Chase Cryogenics (its successful operation demonstrated during the {\it DR-TES} flight), would be held within the cryostat. 

Second, there can be several avenues to further boost the energy and angular resolution of the NV-MMC.
For energy resolutions, the permitted duration of the temperature transient can possibly be further lengthened, leading to even better energy resolution through having the time to collect a greater number of photons.
For instance, lengthening each observation to a 1~second duration from the 10~ms of the projected scenario would improve the photon counts by a factor of 100 and thus the energy resolution by a factor of 10. 
Another potential avenue for optimizing energy resolutions could proceed from replacing the paramagnetic erbium-doped gold absorbers with an alloy (for example, Cu-Ni, as in \cite{2018PhRvX...8a1042W}) having a paramagnetic-ferromagnetic transition at the desired operational temperature. 
As \cite{2018PhRvX...8a1042W} mentions, one could theoretically get as low as $4~\mu$K~Hz$^{-\frac{1}{2}}$ in terms of temperature sensitivity, about 200 times of an improvement over the temperature sensitivity of the NV-MMC, through a more drastic change in magnetization.
One could also use erbium-silver \citep{2018JLTP..193..365K, 2018JLTP..193..435B, 2020PhRvL.125n2503S, 2020JLTP..199.1055K, 2022JLTP..209.1119H}, erbium-copper \citep{2008JLTP..151..363S}, erbium in a semimetal \citep{2000JLTP..121..137E}, or even just palladium \citep{2008JLTP..151..363S} as the absorber material, or even consider utilizing a sensor material reliant on diamagnetism instead of paramagnetism \citep{2012JLTP..167..254B, 2013ITAS...23Q0605S}, to explore other potential optimizations.

For spatial considerations, the optical readout protocol of the NV-MMC can be scaled up to a larger number of absorbers, without needing to worry about the multiplexing required for SQUID readouts---thus improving the field-of-view of future X-ray missions.
Through further reducing the size of the absorbers, the NV-MMC could have potential for further improving its angular resolution as well.

An X-ray detector with improved energy and spatial resolutions could also greatly benefit a wide variety of other applications requiring the precise measurement of X-rays, including medical diagnosis, material characterization, and even nuclear monitoring. 
Even within astrophysics, the wide applicability of MMCs themselves---ranging from proposals for the next {\it NASA} X-ray mission \citep{2022JLTP..209..337D} to detecting beyond-the-Standard-Model particles such as axions \citep{2020arXiv201012076A}---demonstrate the importance of investigating other ways to read out these state-of-the-art MMC detectors. 

\smallskip

\section*{Acknowledgements}
We gratefully acknowledge Kater Murch and James Buckley for valuable discussions. EG and HK acknowledge NASA support from the grants 80NSSC21K1817, 80NSSC22K1883, and 80NSSC24K0205, and thank the McDonnell Center for the Space Sciences at Washington University in St. Louis for their significant support. 
ZL and CZ acknowledge support from NSF QIS 2514391, NSF NRT LinQ 2152221, and the Center for Quantum Leaps at Washington University. 

\paragraph*{Competing Interests:} The authors/co-inventors E.G., Z.L., H.K., and C.Z., at Washington University in St. Louis, have filed for a provisional patent (63/861,426) titled ``Design of a Quantum Diamond Enhanced X-ray Detector''.

\bibliographystyle{naturemag}
\bibliography{Main.bib}

\begin{thebibliography}{100}
\expandafter\ifx\csname url\endcsname\relax
  \def\url#1{\texttt{#1}}\fi
\expandafter\ifx\csname urlprefix\endcsname\relax\def\urlprefix{URL }\fi
\providecommand{\bibinfo}[2]{#2}
\providecommand{\eprint}[2][]{\url{#2}}

\bibitem{2013heai.book.....C}
\bibinfo{author}{{Courvoisier}, T. J.~L.}
\newblock \emph{\bibinfo{title}{{High Energy Astrophysics: An Introduction}}}
  (\bibinfo{year}{2013}).

\bibitem{2022hxga.book.....B}
\bibinfo{author}{{Bambi}, C.} \& \bibinfo{author}{{Sangangelo}, A.}
\newblock \emph{\bibinfo{title}{{Handbook of X-ray and Gamma-ray
  Astrophysics}}} (\bibinfo{year}{2022}).

\bibitem{2017ApJ...848L..12A}
\bibinfo{author}{{Abbott}, B.~P.} \emph{et~al.}
\newblock \bibinfo{title}{{Multi-messenger Observations of a Binary Neutron
  Star Merger}} \textbf{\bibinfo{volume}{848}}, \bibinfo{pages}{L12}
  (\bibinfo{year}{2017}).
\newblock \eprint{1710.05833}.

\bibitem{2022hxga.book..104S}
\bibinfo{author}{{Stratta}, G.} \& \bibinfo{author}{{Santangelo}, A.}
\newblock \bibinfo{title}{{X- and Gamma-Ray Astrophysics in the Era of
  Multi-messenger Astronomy}}.
\newblock In \emph{\bibinfo{booktitle}{Handbook of X-ray and Gamma-ray
  Astrophysics}}, \bibinfo{pages}{104} (\bibinfo{year}{2022}).

\bibitem{2022arXiv221005322B}
\bibinfo{author}{{Bambi}, C.}
\newblock \bibinfo{title}{{Testing Gravity with Black Hole X-Ray Data}}.
\newblock \emph{\bibinfo{journal}{arXiv e-prints}}
  \bibinfo{pages}{arXiv:2210.05322} (\bibinfo{year}{2022}).
\newblock \eprint{2210.05322}.

\bibitem{2018GReGr..50..100K}
\bibinfo{author}{{Krawczynski}, H.}
\newblock \bibinfo{title}{{Difficulties of quantitative tests of the
  Kerr-hypothesis with X-ray observations of mass accreting black holes}}.
\newblock \emph{\bibinfo{journal}{General Relativity and Gravitation}}
  \textbf{\bibinfo{volume}{50}}, \bibinfo{pages}{100} (\bibinfo{year}{2018}).
\newblock \eprint{1806.10347}.

\bibitem{1999astro.ph.12097W}
\bibinfo{author}{{Weisskopf}, M.~C.}
\newblock \bibinfo{title}{{The Chandra X-Ray Observatory (CXO): An Overview}}.
\newblock \emph{\bibinfo{journal}{arXiv e-prints}}
  \bibinfo{pages}{astro--ph/9912097} (\bibinfo{year}{1999}).
\newblock \eprint{astro-ph/9912097}.

\bibitem{2000SPIE.4012....2W}
\bibinfo{author}{{Weisskopf}, M.~C.}, \bibinfo{author}{{Tananbaum}, H.~D.},
  \bibinfo{author}{{Van Speybroeck}, L.~P.} \& \bibinfo{author}{{O'Dell},
  S.~L.}
\newblock \bibinfo{title}{{Chandra X-ray Observatory (CXO): overview}}.
\newblock In \bibinfo{editor}{{Truemper}, J.~E.} \&
  \bibinfo{editor}{{Aschenbach}, B.} (eds.) \emph{\bibinfo{booktitle}{X-Ray
  Optics, Instruments, and Missions III}}, vol. \bibinfo{volume}{4012} of
  \emph{\bibinfo{series}{Society of Photo-Optical Instrumentation Engineers
  (SPIE) Conference Series}}, \bibinfo{pages}{2--16} (\bibinfo{year}{2000}).
\newblock \eprint{astro-ph/0004127}.

\bibitem{ChandraCalWebsiteAcis}
\bibinfo{title}{Acis instrument information}.
\newblock \bibinfo{howpublished}{\url{https://cxc.harvard.edu/cal/Acis/}}.
\newblock \bibinfo{note}{Accessed: 2025-03-15}.

\bibitem{ChandraCalWebsiteHrc}
\bibinfo{title}{Hrc instrument information}.
\newblock
  \bibinfo{howpublished}{\url{https://cxc.harvard.edu/cal/Hrc/index.html}}.
\newblock \bibinfo{note}{Accessed: 2025-03-15}.

\bibitem{2002PASP..114....1W}
\bibinfo{author}{{Weisskopf}, M.~C.} \emph{et~al.}
\newblock \bibinfo{title}{{An Overview of the Performance and Scientific
  Results from the Chandra X-Ray Observatory}} \textbf{\bibinfo{volume}{114}},
  \bibinfo{pages}{1--24} (\bibinfo{year}{2002}).
\newblock \eprint{astro-ph/0110308}.

\bibitem{2022hxga.book...84W}
\bibinfo{author}{{Wilkes}, B.~J.} \& \bibinfo{author}{{Tananbaum}, H.}
\newblock \bibinfo{title}{{The Chandra X-ray Observatory}}.
\newblock In \emph{\bibinfo{booktitle}{Handbook of X-ray and Gamma-ray
  Astrophysics. Edited by Cosimo Bambi and Andrea Santangelo}},
  \bibinfo{pages}{84} (\bibinfo{year}{2022}).

\bibitem{2020arXiv200304962X}
\bibinfo{author}{{XRISM Science Team}}.
\newblock \bibinfo{title}{{Science with the X-ray Imaging and Spectroscopy
  Mission (XRISM)}}.
\newblock \emph{\bibinfo{journal}{arXiv e-prints}}
  \bibinfo{pages}{arXiv:2003.04962} (\bibinfo{year}{2020}).
\newblock \eprint{2003.04962}.

\bibitem{2022arXiv220205399X}
\bibinfo{author}{{XRISM Science Team}}.
\newblock \bibinfo{title}{{XRISM Quick Reference}}.
\newblock \emph{\bibinfo{journal}{arXiv e-prints}}
  \bibinfo{pages}{arXiv:2202.05399} (\bibinfo{year}{2022}).
\newblock \eprint{2202.05399}.

\bibitem{2022SPIE12181E..1TM}
\bibinfo{author}{{Mori}, K.} \emph{et~al.}
\newblock \bibinfo{title}{{Xtend, the soft x-ray imaging telescope for the
  X-Ray Imaging and Spectroscopy Mission (XRISM)}}.
\newblock In \bibinfo{editor}{{den Herder}, J.-W.~A.},
  \bibinfo{editor}{{Nikzad}, S.} \& \bibinfo{editor}{{Nakazawa}, K.} (eds.)
  \emph{\bibinfo{booktitle}{Space Telescopes and Instrumentation 2022:
  Ultraviolet to Gamma Ray}}, vol. \bibinfo{volume}{12181} of
  \emph{\bibinfo{series}{Society of Photo-Optical Instrumentation Engineers
  (SPIE) Conference Series}}, \bibinfo{pages}{121811T} (\bibinfo{year}{2022}).
\newblock \eprint{2303.07575}.

\bibitem{2022SPIE12181E..5ZM}
\bibinfo{author}{{Mizumoto}, M.} \emph{et~al.}
\newblock \bibinfo{title}{{High count rate effects in event processing for
  XRISM/Resolve x-ray microcalorimeter}}.
\newblock In \bibinfo{editor}{{den Herder}, J.-W.~A.},
  \bibinfo{editor}{{Nikzad}, S.} \& \bibinfo{editor}{{Nakazawa}, K.} (eds.)
  \emph{\bibinfo{booktitle}{Space Telescopes and Instrumentation 2022:
  Ultraviolet to Gamma Ray}}, vol. \bibinfo{volume}{12181} of
  \emph{\bibinfo{series}{Society of Photo-Optical Instrumentation Engineers
  (SPIE) Conference Series}}, \bibinfo{pages}{121815Z} (\bibinfo{year}{2022}).
\newblock \eprint{2312.15588}.

\bibitem{2023arXiv230301642S}
\bibinfo{author}{{Sato}, K.}, \bibinfo{author}{{Uchida}, Y.} \&
  \bibinfo{author}{{Ishikawa}, K.}
\newblock \bibinfo{title}{{Hitomi/XRISM micro-calorimeter}}.
\newblock \emph{\bibinfo{journal}{arXiv e-prints}}
  \bibinfo{pages}{arXiv:2303.01642} (\bibinfo{year}{2023}).
\newblock \eprint{2303.01642}.

\bibitem{2005cpd..book...63I}
\bibinfo{author}{{Irwin}, K.~D.} \& \bibinfo{author}{{Hilton}, G.~C.}
\newblock \bibinfo{title}{{Transition-Edge Sensors}}.
\newblock In \bibinfo{editor}{{Enss}, C.} (ed.)
  \emph{\bibinfo{booktitle}{Cryogenic Particle Detection}},
  vol.~\bibinfo{volume}{99}, \bibinfo{pages}{63} (\bibinfo{year}{2005}).

\bibitem{2005ApPhL..87s4103U}
\bibinfo{author}{{Ullom}, J.~N.} \emph{et~al.}
\newblock \bibinfo{title}{{Optimized transition-edge x-ray microcalorimeter
  with 2.4 eV energy resolution at 5.9 keV}}.
\newblock \emph{\bibinfo{journal}{Applied Physics Letters}}
  \textbf{\bibinfo{volume}{87}}, \bibinfo{pages}{194103}
  (\bibinfo{year}{2005}).

\bibitem{2009ITAS...19..451S}
\bibinfo{author}{{Smith}, S.~J.} \emph{et~al.}
\newblock \bibinfo{title}{{Development of Position-Sensitive Transition-Edge
  Sensor X-Ray Detectors}}.
\newblock \emph{\bibinfo{journal}{IEEE Transactions on Applied
  Superconductivity}} \textbf{\bibinfo{volume}{19}}, \bibinfo{pages}{451--455}
  (\bibinfo{year}{2009}).

\bibitem{2009ITNS...56.2299B}
\bibinfo{author}{{Bacrania}, M.~K.} \emph{et~al.}
\newblock \bibinfo{title}{{Large-Area Microcalorimeter Detectors for
  Ultra-High-Resolution X-Ray and Gamma-Ray Spectroscopy}}.
\newblock \emph{\bibinfo{journal}{IEEE Transactions on Nuclear Science}}
  \textbf{\bibinfo{volume}{56}}, \bibinfo{pages}{2299--2302}
  (\bibinfo{year}{2009}).

\bibitem{2012RScI...83i3113B}
\bibinfo{author}{{Bennett}, D.~A.} \emph{et~al.}
\newblock \bibinfo{title}{{A high resolution gamma-ray spectrometer based on
  superconducting microcalorimeters}}.
\newblock \emph{\bibinfo{journal}{Review of Scientific Instruments}}
  \textbf{\bibinfo{volume}{83}}, \bibinfo{pages}{093113--093113--14}
  (\bibinfo{year}{2012}).

\bibitem{2015ApPhL.107v3503L}
\bibinfo{author}{{Lee}, S.~J.} \emph{et~al.}
\newblock \bibinfo{title}{{Fine pitch transition-edge sensor X-ray
  microcalorimeters with sub-eV energy resolution at 1.5 keV}}.
\newblock \emph{\bibinfo{journal}{Applied Physics Letters}}
  \textbf{\bibinfo{volume}{107}}, \bibinfo{pages}{223503}
  (\bibinfo{year}{2015}).

\bibitem{2020JLTP..199..949S}
\bibinfo{author}{{Sakai}, K.} \emph{et~al.}
\newblock \bibinfo{title}{{Demonstration of Fine-Pitch High-Resolution X-ray
  Transition-Edge Sensor Microcalorimeters Optimized for Energies below 1
  keV}}.
\newblock \emph{\bibinfo{journal}{Journal of Low Temperature Physics}}
  \textbf{\bibinfo{volume}{199}}, \bibinfo{pages}{949--954}
  (\bibinfo{year}{2020}).

\bibitem{1999ApPhL..74.4043C}
\bibinfo{author}{{Chervenak}, J.~A.} \emph{et~al.}
\newblock \bibinfo{title}{{Superconducting multiplexer for arrays of transition
  edge sensors}}.
\newblock \emph{\bibinfo{journal}{Applied Physics Letters}}
  \textbf{\bibinfo{volume}{74}}, \bibinfo{pages}{4043} (\bibinfo{year}{1999}).

\bibitem{2001ApPhL..78..371Y}
\bibinfo{author}{{Yoon}, J.} \emph{et~al.}
\newblock \bibinfo{title}{{Single superconducting quantum interference device
  multiplexer for arrays of low-temperature sensors}}.
\newblock \emph{\bibinfo{journal}{Applied Physics Letters}}
  \textbf{\bibinfo{volume}{78}}, \bibinfo{pages}{371} (\bibinfo{year}{2001}).

\bibitem{2010SuScT..23c4004I}
\bibinfo{author}{{Irwin}, K.~D.} \emph{et~al.}
\newblock \bibinfo{title}{{Code-division multiplexing of superconducting
  transition-edge sensor arrays}}.
\newblock \emph{\bibinfo{journal}{Superconductor Science Technology}}
  \textbf{\bibinfo{volume}{23}}, \bibinfo{pages}{034004}
  (\bibinfo{year}{2010}).

\bibitem{2004ApPhL..85.2107I}
\bibinfo{author}{{Irwin}, K.~D.} \& \bibinfo{author}{{Lehnert}, K.~W.}
\newblock \bibinfo{title}{{Microwave SQUID multiplexer}}.
\newblock \emph{\bibinfo{journal}{Applied Physics Letters}}
  \textbf{\bibinfo{volume}{85}}, \bibinfo{pages}{2107} (\bibinfo{year}{2004}).

\bibitem{2008ApPhL..92b3514M}
\bibinfo{author}{{Mates}, J.~A.~B.}, \bibinfo{author}{{Hilton}, G.~C.},
  \bibinfo{author}{{Irwin}, K.~D.}, \bibinfo{author}{{Vale}, L.~R.} \&
  \bibinfo{author}{{Lehnert}, K.~W.}
\newblock \bibinfo{title}{{Demonstration of a multiplexer of dissipationless
  superconducting quantum interference devices}}.
\newblock \emph{\bibinfo{journal}{Applied Physics Letters}}
  \textbf{\bibinfo{volume}{92}}, \bibinfo{pages}{023514}
  (\bibinfo{year}{2008}).

\bibitem{2011PhDT........55M}
\bibinfo{author}{{Mates}, J. A.~B.}
\newblock \emph{\bibinfo{title}{{The Microwave SQUID Multiplexer}}}.
\newblock Ph.D. thesis, \bibinfo{school}{University of Colorado, Boulder}
  (\bibinfo{year}{2011}).

\bibitem{2012JLTP..167..707M}
\bibinfo{author}{{Mates}, J.~A.~B.} \emph{et~al.}
\newblock \bibinfo{title}{{Flux-Ramp Modulation for SQUID Multiplexing}}.
\newblock \emph{\bibinfo{journal}{Journal of Low Temperature Physics}}
  \textbf{\bibinfo{volume}{167}}, \bibinfo{pages}{707--712}
  (\bibinfo{year}{2012}).

\bibitem{2016ApPhL.109k2604M}
\bibinfo{author}{{Morgan}, K.~M.} \emph{et~al.}
\newblock \bibinfo{title}{{Code-division-multiplexed readout of large arrays of
  TES microcalorimeters}}.
\newblock \emph{\bibinfo{journal}{Applied Physics Letters}}
  \textbf{\bibinfo{volume}{109}}, \bibinfo{pages}{112604}
  (\bibinfo{year}{2016}).

\bibitem{2017ApPhL.111f2601M}
\bibinfo{author}{{Mates}, J.~A.~B.} \emph{et~al.}
\newblock \bibinfo{title}{{Simultaneous readout of 128 X-ray and gamma-ray
  transition-edge microcalorimeters using microwave SQUID multiplexing}}.
\newblock \emph{\bibinfo{journal}{Applied Physics Letters}}
  \textbf{\bibinfo{volume}{111}}, \bibinfo{pages}{062601}
  (\bibinfo{year}{2017}).

\bibitem{2018JLTP..193..485G}
\bibinfo{author}{{Gard}, J.~D.} \emph{et~al.}
\newblock \bibinfo{title}{{A Scalable Readout for Microwave SQUID Multiplexing
  of Transition-Edge Sensors}}.
\newblock \emph{\bibinfo{journal}{Journal of Low Temperature Physics}}
  \textbf{\bibinfo{volume}{193}}, \bibinfo{pages}{485--497}
  (\bibinfo{year}{2018}).

\bibitem{2019ITAS...2904472D}
\bibinfo{author}{{Durkin}, M.} \emph{et~al.}
\newblock \bibinfo{title}{{Demonstration of Athena X-IFU Compatible 40-Row
  Time-Division-Multiplexed Readout}}.
\newblock \emph{\bibinfo{journal}{IEEE Transactions on Applied
  Superconductivity}} \textbf{\bibinfo{volume}{29}}, \bibinfo{pages}{2904472}
  (\bibinfo{year}{2019}).

\bibitem{2019JATIS...5b1007B}
\bibinfo{author}{{Bennett}, D.~A.} \emph{et~al.}
\newblock \bibinfo{title}{{Microwave SQUID multiplexing for the Lynx x-ray
  microcalorimeter}}.
\newblock \emph{\bibinfo{journal}{Journal of Astronomical Telescopes,
  Instruments, and Systems}} \textbf{\bibinfo{volume}{5}},
  \bibinfo{pages}{021007} (\bibinfo{year}{2019}).

\bibitem{2019JATIS...5b1008S}
\bibinfo{author}{{Smith}, S.~J.} \emph{et~al.}
\newblock \bibinfo{title}{{Multiabsorber transition-edge sensors for x-ray
  astronomy}}.
\newblock \emph{\bibinfo{journal}{Journal of Astronomical Telescopes,
  Instruments, and Systems}} \textbf{\bibinfo{volume}{5}},
  \bibinfo{pages}{021008} (\bibinfo{year}{2019}).

\bibitem{2020JLTP..199..330S}
\bibinfo{author}{{Smith}, S.~J.} \emph{et~al.}
\newblock \bibinfo{title}{{Toward 100,000-Pixel Microcalorimeter Arrays Using
  Multi-absorber Transition-Edge Sensors}}.
\newblock \emph{\bibinfo{journal}{Journal of Low Temperature Physics}}
  \textbf{\bibinfo{volume}{199}}, \bibinfo{pages}{330--338}
  (\bibinfo{year}{2020}).
\newblock \eprint{1908.02687}.

\bibitem{2023JATIS...9d1006W}
\bibinfo{author}{{Wakeham}, N.} \emph{et~al.}
\newblock \bibinfo{title}{{Characterization of a hybrid array of single and
  multi-absorber transition-edge sensor microcalorimeters for the Line Emission
  Mapper}}.
\newblock \emph{\bibinfo{journal}{Journal of Astronomical Telescopes,
  Instruments, and Systems}} \textbf{\bibinfo{volume}{9}},
  \bibinfo{pages}{041006} (\bibinfo{year}{2023}).

\bibitem{2021APh...12602529A}
\bibinfo{author}{{Abarr}, Q.} \emph{et~al.}
\newblock \bibinfo{title}{{XL-Calibur - a second-generation balloon-borne hard
  X-ray polarimetry mission}}.
\newblock \emph{\bibinfo{journal}{Astroparticle Physics}}
  \textbf{\bibinfo{volume}{126}}, \bibinfo{pages}{102529}
  (\bibinfo{year}{2021}).
\newblock \eprint{2010.10608}.

\bibitem{1993JLTP...93..709B}
\bibinfo{author}{{Bandler}, S.~R.} \emph{et~al.}
\newblock \bibinfo{title}{{Metallic magnetic bolometers for particle
  detection}}.
\newblock \emph{\bibinfo{journal}{Journal of Low Temperature Physics}}
  \textbf{\bibinfo{volume}{93}}, \bibinfo{pages}{709--714}
  (\bibinfo{year}{1993}).

\bibitem{1999PhyB..263..604A}
\bibinfo{author}{{Adams}, J.~S.} \emph{et~al.}
\newblock \bibinfo{title}{{Particle detection using cryogenic magnetic
  calorimeters}}.
\newblock \emph{\bibinfo{journal}{Physica B Condensed Matter}}
  \textbf{\bibinfo{volume}{263}}, \bibinfo{pages}{604--606}
  (\bibinfo{year}{1999}).

\bibitem{2000JLTP..118....7F}
\bibinfo{author}{{Fleischmann}, A.} \emph{et~al.}
\newblock \emph{\bibinfo{journal}{Journal of Low Temperature Physics}}
  \textbf{\bibinfo{volume}{118}}, \bibinfo{pages}{7--21}
  (\bibinfo{year}{2000}).

\bibitem{2000JLTP..121..137E}
\bibinfo{author}{{Enss}, C.} \emph{et~al.}
\newblock \emph{\bibinfo{journal}{Journal of Low Temperature Physics}}
  \textbf{\bibinfo{volume}{121}}, \bibinfo{pages}{137--176}
  (\bibinfo{year}{2000}).

\bibitem{2000NIMPA.444..100F}
\bibinfo{author}{{Fleischmann}, A.} \emph{et~al.}
\newblock \bibinfo{title}{{The sensitivity of magnetic calorimeters with large
  heat capacity}}.
\newblock \emph{\bibinfo{journal}{Nuclear Instruments and Methods in Physics
  Research A}} \textbf{\bibinfo{volume}{444}}, \bibinfo{pages}{100--103}
  (\bibinfo{year}{2000}).

\bibitem{2000NIMPA.444..211S}
\bibinfo{author}{{Sch{\"o}nefeld}, J.} \emph{et~al.}
\newblock \bibinfo{title}{{X-ray detection using magnetic calorimeters}}.
\newblock \emph{\bibinfo{journal}{Nuclear Instruments and Methods in Physics
  Research A}} \textbf{\bibinfo{volume}{444}}, \bibinfo{pages}{211--213}
  (\bibinfo{year}{2000}).

\bibitem{2000SPIE.4140..445A}
\bibinfo{author}{{Adams}, J.~S.} \emph{et~al.}
\newblock \bibinfo{title}{{Magnetic calorimeters for x-ray and gamma-ray
  detection}}.
\newblock In \bibinfo{editor}{{Flanagan}, K.~A.} \&
  \bibinfo{editor}{{Siegmund}, O.~H.} (eds.) \emph{\bibinfo{booktitle}{X-Ray
  and Gamma-Ray Instrumentation for Astronomy XI}}, vol. \bibinfo{volume}{4140}
  of \emph{\bibinfo{series}{Society of Photo-Optical Instrumentation Engineers
  (SPIE) Conference Series}}, \bibinfo{pages}{445--451} (\bibinfo{year}{2000}).

\bibitem{2003PhyB..329.1594F}
\bibinfo{author}{{Fleischmann}, A.}, \bibinfo{author}{{Daniyarov}, T.},
  \bibinfo{author}{{Rotzinger}, H.}, \bibinfo{author}{{Enss}, C.} \&
  \bibinfo{author}{{Seidel}, G.}
\newblock \bibinfo{title}{{Magnetic calorimeters for high resolution X-ray
  spectroscopy}}.
\newblock \emph{\bibinfo{journal}{Physica B Condensed Matter}}
  \textbf{\bibinfo{volume}{329}}, \bibinfo{pages}{1594--1595}
  (\bibinfo{year}{2003}).

\bibitem{2003RScI...74.3947F}
\bibinfo{author}{{Fleischmann}, A.} \emph{et~al.}
\newblock \bibinfo{title}{{Magnetic calorimeters for high resolution x-ray
  spectroscopy}}.
\newblock \emph{\bibinfo{journal}{Review of Scientific Instruments}}
  \textbf{\bibinfo{volume}{74}}, \bibinfo{pages}{3947--3954}
  (\bibinfo{year}{2003}).

\bibitem{2003SuScT..16.1404Z}
\bibinfo{author}{{Zakosarenko}, V.} \emph{et~al.}
\newblock \bibinfo{title}{{SQUID gradiometer for ultra-low temperature magnetic
  micro-calorimeter}}.
\newblock \emph{\bibinfo{journal}{Superconductor Science Technology}}
  \textbf{\bibinfo{volume}{16}}, \bibinfo{pages}{1404--1407}
  (\bibinfo{year}{2003}).

\bibitem{2004NIMPA.520...27F}
\bibinfo{author}{{Fleischmann}, A.} \emph{et~al.}
\newblock \bibinfo{title}{{Metallic magnetic calorimeters (MMC): detectors for
  high-resolution X-ray spectroscopy}}.
\newblock \emph{\bibinfo{journal}{Nuclear Instruments and Methods in Physics
  Research A}} \textbf{\bibinfo{volume}{520}}, \bibinfo{pages}{27--31}
  (\bibinfo{year}{2004}).

\bibitem{2004NIMPA.520...52Z}
\bibinfo{author}{{Zink}, B.~L.} \emph{et~al.}
\newblock \bibinfo{title}{{Lithographically patterned magnetic calorimeter
  X-ray detectors with integrated SQUID readout}}.
\newblock \emph{\bibinfo{journal}{Nuclear Instruments and Methods in Physics
  Research A}} \textbf{\bibinfo{volume}{520}}, \bibinfo{pages}{52--55}
  (\bibinfo{year}{2004}).

\bibitem{2004PSSCR...1.2824F}
\bibinfo{author}{{Fleischmann}, A.} \emph{et~al.}
\newblock \bibinfo{title}{{Metallic magnetic microcalorimeters: Energy
  dispersive X-ray detectors with high spectral resolving power}}.
\newblock \emph{\bibinfo{journal}{Physica Status Solidi C Current Topics}}
  \textbf{\bibinfo{volume}{1}}, \bibinfo{pages}{2824--2827}
  (\bibinfo{year}{2004}).

\bibitem{2004SPIE.5501..346B}
\bibinfo{author}{{Bandler}, S.~R.}
\newblock \bibinfo{title}{{Development of magnetic microcalorimeters for X-ray
  astronomy}}.
\newblock In \bibinfo{editor}{{Holland}, A.~D.} (ed.)
  \emph{\bibinfo{booktitle}{High-Energy Detectors in Astronomy}}, vol.
  \bibinfo{volume}{5501} of \emph{\bibinfo{series}{Society of Photo-Optical
  Instrumentation Engineers (SPIE) Conference Series}},
  \bibinfo{pages}{346--356} (\bibinfo{year}{2004}).

\bibitem{2005cpd..book..151F}
\bibinfo{author}{{Fleischmann}, A.}, \bibinfo{author}{{Enss}, C.} \&
  \bibinfo{author}{{Seidel}, G.~M.}
\newblock \emph{\bibinfo{title}{{Metallic Magnetic Calorimeters}}},
  vol.~\bibinfo{volume}{99}, \bibinfo{pages}{151} (\bibinfo{year}{2005}).

\bibitem{2005ITAS...15..773S}
\bibinfo{author}{{Stolz}, R.} \emph{et~al.}
\newblock \bibinfo{title}{{SQUID-Gradiometers for Arrays of Integrated Low
  Temperature Magnetic Micro-Calorimeters}}.
\newblock \emph{\bibinfo{journal}{IEEE Transactions on Applied
  Superconductivity}} \textbf{\bibinfo{volume}{15}}, \bibinfo{pages}{773--776}
  (\bibinfo{year}{2005}).

\bibitem{2006JAP....99hB303Z}
\bibinfo{author}{{Zink}, B.~L.}, \bibinfo{author}{{Irwin}, K.~D.},
  \bibinfo{author}{{Hilton}, G.~C.}, \bibinfo{author}{{Ullom}, J.~N.} \&
  \bibinfo{author}{{Pappas}, D.~P.}
\newblock \bibinfo{title}{{Erbium-doped gold sensor films for magnetic
  microcalorimeter x-ray detectors}}.
\newblock \emph{\bibinfo{journal}{Journal of Applied Physics}}
  \textbf{\bibinfo{volume}{99}}, \bibinfo{pages}{08B303}
  (\bibinfo{year}{2006}).

\bibitem{2008JLTP..151..345S}
\bibinfo{author}{{Sato}, K.}, \bibinfo{author}{{Tsuchiya}, A.},
  \bibinfo{author}{{Oshima}, T.}, \bibinfo{author}{{Yamasaki}, N.~Y.} \&
  \bibinfo{author}{{Morooka}, T.}
\newblock \bibinfo{title}{{Development of Low Temperature SQUID Gradiometer
  Array for Metallic Magnetic Microcalorimeters}}.
\newblock \emph{\bibinfo{journal}{Journal of Low Temperature Physics}}
  \textbf{\bibinfo{volume}{151}}, \bibinfo{pages}{345--350}
  (\bibinfo{year}{2008}).

\bibitem{2008JLTP..151..357H}
\bibinfo{author}{{Hsieh}, W.~T.} \emph{et~al.}
\newblock \bibinfo{title}{{Fabrication of Metallic Magnetic Calorimeter X-ray
  Detector Arrays}}.
\newblock \emph{\bibinfo{journal}{Journal of Low Temperature Physics}}
  \textbf{\bibinfo{volume}{151}}, \bibinfo{pages}{357--362}
  (\bibinfo{year}{2008}).

\bibitem{2008JLTP..151..363S}
\bibinfo{author}{{Sultan}, R.} \emph{et~al.}
\newblock \bibinfo{title}{{Design, Fabrication, and Multiplexing of Magnetic
  Calorimeter X-ray Detectors with High-Efficiency SQUID Readout}}.
\newblock \emph{\bibinfo{journal}{Journal of Low Temperature Physics}}
  \textbf{\bibinfo{volume}{151}}, \bibinfo{pages}{363--368}
  (\bibinfo{year}{2008}).

\bibitem{2008JLTP..151..337B}
\bibinfo{author}{{Burck}, A.} \emph{et~al.}
\newblock \bibinfo{title}{{Microstructured Magnetic Calorimeter with
  Meander-Shaped Pickup Coil}}.
\newblock \emph{\bibinfo{journal}{Journal of Low Temperature Physics}}
  \textbf{\bibinfo{volume}{151}}, \bibinfo{pages}{337--344}
  (\bibinfo{year}{2008}).

\bibitem{2008SPIE.7021E..1KB}
\bibinfo{author}{{Bandler}, S.~R.} \emph{et~al.}
\newblock \bibinfo{title}{{Micro-fabricated magnetic microcalorimeter
  development for x-ray astronomy}}.
\newblock In \bibinfo{editor}{{Dorn}, D.~A.} \& \bibinfo{editor}{{Holland},
  A.~D.} (eds.) \emph{\bibinfo{booktitle}{High Energy, Optical, and Infrared
  Detectors for Astronomy III}}, vol. \bibinfo{volume}{7021} of
  \emph{\bibinfo{series}{Society of Photo-Optical Instrumentation Engineers
  (SPIE) Conference Series}}, \bibinfo{pages}{70211K} (\bibinfo{year}{2008}).

\bibitem{2009AIPC.1185..571F}
\bibinfo{author}{{Fleischmann}, A.} \emph{et~al.}
\newblock \bibinfo{title}{{Metallic magnetic calorimeters}}.
\newblock In \bibinfo{editor}{{Young}, B.}, \bibinfo{editor}{{Cabrera}, B.} \&
  \bibinfo{editor}{{Miller}, A.} (eds.) \emph{\bibinfo{booktitle}{The
  Thirteenth International Workshop on Low Temperature Detectors - LTD13}},
  vol. \bibinfo{volume}{1185} of \emph{\bibinfo{series}{American Institute of
  Physics Conference Series}}, \bibinfo{pages}{571--578}
  (\bibinfo{year}{2009}).

\bibitem{2009AIPC.1185..579B}
\bibinfo{author}{{Bandler}, S.~R.} \emph{et~al.}
\newblock \bibinfo{title}{{Performance of High-Resolution, Micro-fabricated,
  X-ray Magnetic Calorimeters}}.
\newblock In \bibinfo{editor}{{Young}, B.}, \bibinfo{editor}{{Cabrera}, B.} \&
  \bibinfo{editor}{{Miller}, A.} (eds.) \emph{\bibinfo{booktitle}{The
  Thirteenth International Workshop on Low Temperature Detectors - LTD13}},
  vol. \bibinfo{volume}{1185} of \emph{\bibinfo{series}{American Institute of
  Physics Conference Series}}, \bibinfo{pages}{579--582}
  (\bibinfo{publisher}{AIP}, \bibinfo{year}{2009}).

\bibitem{2009AIPC.1185..583R}
\bibinfo{author}{{Rodrigues}, M.} \emph{et~al.}
\newblock \bibinfo{title}{{Metallic magnetic calorimeters for
  {\ensuremath{\gamma}}-ray spectrometry}}.
\newblock In \bibinfo{editor}{{Young}, B.}, \bibinfo{editor}{{Cabrera}, B.} \&
  \bibinfo{editor}{{Miller}, A.} (eds.) \emph{\bibinfo{booktitle}{The
  Thirteenth International Workshop on Low Temperature Detectors - LTD13}},
  vol. \bibinfo{volume}{1185} of \emph{\bibinfo{series}{American Institute of
  Physics Conference Series}}, \bibinfo{pages}{583--586}
  (\bibinfo{publisher}{AIP}, \bibinfo{year}{2009}).

\bibitem{2009ITAS...19...63F}
\bibinfo{author}{{Fleischmann}, L.} \emph{et~al.}
\newblock \bibinfo{title}{{Metallic Magnetic Calorimeters for X-Ray
  Spectroscopy}}.
\newblock \emph{\bibinfo{journal}{IEEE Transactions on Applied
  Superconductivity}} \textbf{\bibinfo{volume}{19}}, \bibinfo{pages}{63--68}
  (\bibinfo{year}{2009}).

\bibitem{2009JPhCS.150a2013F}
\bibinfo{author}{{Fleischmann}, L.} \emph{et~al.}
\newblock \bibinfo{title}{{Physics and applications of metallic magnetic
  calorimeters for particle detection}}.
\newblock In \emph{\bibinfo{booktitle}{Journal of Physics Conference Series}},
  vol. \bibinfo{volume}{150} of \emph{\bibinfo{series}{Journal of Physics
  Conference Series}}, \bibinfo{pages}{012013} (\bibinfo{publisher}{IOP},
  \bibinfo{year}{2009}).

\bibitem{2012JLTP..167..254B}
\bibinfo{author}{{Bandler}, S.~R.} \emph{et~al.}
\newblock \bibinfo{title}{{Magnetically Coupled Microcalorimeters}}.
\newblock \emph{\bibinfo{journal}{Journal of Low Temperature Physics}}
  \textbf{\bibinfo{volume}{167}}, \bibinfo{pages}{254--268}
  (\bibinfo{year}{2012}).

\bibitem{2012JLTP..167..269P}
\bibinfo{author}{{Pies}, C.} \emph{et~al.}
\newblock \bibinfo{title}{{maXs: Microcalorimeter Arrays for High-Resolution
  X-Ray Spectroscopy at GSI/FAIR}}.
\newblock \emph{\bibinfo{journal}{Journal of Low Temperature Physics}}
  \textbf{\bibinfo{volume}{167}}, \bibinfo{pages}{269--279}
  (\bibinfo{year}{2012}).

\bibitem{2014JLTP..176..617P}
\bibinfo{author}{{Porst}, J.~P.} \emph{et~al.}
\newblock \bibinfo{title}{{Characterization and Performance of Magnetic
  Calorimeters for Applications in X-ray Spectroscopy}}.
\newblock \emph{\bibinfo{journal}{Journal of Low Temperature Physics}}
  \textbf{\bibinfo{volume}{176}}, \bibinfo{pages}{617--623}
  (\bibinfo{year}{2014}).

\bibitem{2016JLTP..184..108L}
\bibinfo{author}{{Le}, L.~N.}, \bibinfo{author}{{Hummatov}, R.},
  \bibinfo{author}{{Hall}, J.~A.}, \bibinfo{author}{{Cantor}, R.~C.} \&
  \bibinfo{author}{{Boyd}, S.~T.~P.}
\newblock \bibinfo{title}{{Development of Magnetic Microcalorimeters for
  Gamma-Ray Spectroscopy}}.
\newblock \emph{\bibinfo{journal}{Journal of Low Temperature Physics}}
  \textbf{\bibinfo{volume}{184}}, \bibinfo{pages}{108--113}
  (\bibinfo{year}{2016}).

\bibitem{2016JLTP..184..344K}
\bibinfo{author}{{Kempf}, S.}, \bibinfo{author}{{Ferring}, A.},
  \bibinfo{author}{{Fleischmann}, A.}, \bibinfo{author}{{Wegner}, M.} \&
  \bibinfo{author}{{Enss}, C.}
\newblock \bibinfo{title}{{Metallic Magnetic Calorimeters with On-Chip dc-SQUID
  Readout}}.
\newblock \emph{\bibinfo{journal}{Journal of Low Temperature Physics}}
  \textbf{\bibinfo{volume}{184}}, \bibinfo{pages}{344--350}
  (\bibinfo{year}{2016}).

\bibitem{2017AIPA....7a5007K}
\bibinfo{author}{{Kempf}, S.} \emph{et~al.}
\newblock \bibinfo{title}{{Demonstration of a scalable frequency-domain readout
  of metallic magnetic calorimeters by means of a microwave SQUID
  multiplexer}}.
\newblock \emph{\bibinfo{journal}{AIP Advances}} \textbf{\bibinfo{volume}{7}},
  \bibinfo{pages}{015007} (\bibinfo{year}{2017}).

\bibitem{2017SuScT..30f5002K}
\bibinfo{author}{{Kempf}, S.} \emph{et~al.}
\newblock \bibinfo{title}{{Design, fabrication and characterization of a 64
  pixel metallic magnetic calorimeter array with integrated, on-chip microwave
  SQUID multiplexer}}.
\newblock \emph{\bibinfo{journal}{Superconductor Science Technology}}
  \textbf{\bibinfo{volume}{30}}, \bibinfo{pages}{065002}
  (\bibinfo{year}{2017}).

\bibitem{2018JLTP..193..365K}
\bibinfo{author}{{Kempf}, S.}, \bibinfo{author}{{Fleischmann}, A.},
  \bibinfo{author}{{Gastaldo}, L.} \& \bibinfo{author}{{Enss}, C.}
\newblock \bibinfo{title}{{Physics and Applications of Metallic Magnetic
  Calorimeters}}.
\newblock \emph{\bibinfo{journal}{Journal of Low Temperature Physics}}
  \textbf{\bibinfo{volume}{193}}, \bibinfo{pages}{365--379}
  (\bibinfo{year}{2018}).

\bibitem{2018JLTP..193..462W}
\bibinfo{author}{{Wegner}, M.} \emph{et~al.}
\newblock \bibinfo{title}{{Microwave SQUID Multiplexing of Metallic Magnetic
  Calorimeters: Status of Multiplexer Performance and Room-Temperature Readout
  Electronics Development}}.
\newblock \emph{\bibinfo{journal}{Journal of Low Temperature Physics}}
  \textbf{\bibinfo{volume}{193}}, \bibinfo{pages}{462--475}
  (\bibinfo{year}{2018}).

\bibitem{2018JLTP..193..668S}
\bibinfo{author}{{Stevenson}, T.~R.} \emph{et~al.}
\newblock \bibinfo{title}{{Magnetic Calorimeter Arrays with High Sensor
  Inductance and Dense Wiring}}.
\newblock \emph{\bibinfo{journal}{Journal of Low Temperature Physics}}
  \textbf{\bibinfo{volume}{193}}, \bibinfo{pages}{668--674}
  (\bibinfo{year}{2018}).

\bibitem{2019ITAS...2908143Y}
\bibinfo{author}{{Yoon}, W.} \emph{et~al.}
\newblock \bibinfo{title}{{Design and Performance of a Prototype Magnetic
  Calorimeter Array for the Lynx X-Ray Microcalorimeter}}.
\newblock \emph{\bibinfo{journal}{IEEE Transactions on Applied
  Superconductivity}} \textbf{\bibinfo{volume}{29}}, \bibinfo{pages}{2908143}
  (\bibinfo{year}{2019}).

\bibitem{2019JATIS...5b1009S}
\bibinfo{author}{{Stevenson}, T.~R.} \emph{et~al.}
\newblock \bibinfo{title}{{Magnetic calorimeter option for the Lynx x-ray
  microcalorimeter}}.
\newblock \emph{\bibinfo{journal}{Journal of Astronomical Telescopes,
  Instruments, and Systems}} \textbf{\bibinfo{volume}{5}},
  \bibinfo{pages}{021009} (\bibinfo{year}{2019}).

\bibitem{2020JLTP..199..916Y}
\bibinfo{author}{{Yoon}, W.} \emph{et~al.}
\newblock \bibinfo{title}{{Position-Sensitive Magnetic Calorimeters for Lynx}}.
\newblock \emph{\bibinfo{journal}{Journal of Low Temperature Physics}}
  \textbf{\bibinfo{volume}{199}}, \bibinfo{pages}{916--922}
  (\bibinfo{year}{2020}).

\bibitem{2021JInst..16P8003M}
\bibinfo{author}{{Mantegazzini}, F.} \emph{et~al.}
\newblock \bibinfo{title}{{Multichannel read-out for arrays of metallic
  magnetic calorimeters}}.
\newblock \emph{\bibinfo{journal}{Journal of Instrumentation}}
  \textbf{\bibinfo{volume}{16}}, \bibinfo{pages}{P08003}
  (\bibinfo{year}{2021}).
\newblock \eprint{2102.11100}.

\bibitem{2022JLTP..209..337D}
\bibinfo{author}{{Devasia}, A.~M.}, \bibinfo{author}{{Bandler}, S.~R.},
  \bibinfo{author}{{Ryu}, K.}, \bibinfo{author}{{Stevenson}, T.~R.} \&
  \bibinfo{author}{{Yoon}, W.}
\newblock \bibinfo{title}{{Large-Scale Magnetic Microcalorimeter Arrays for the
  Lynx X-Ray Microcalorimeter}}.
\newblock \emph{\bibinfo{journal}{Journal of Low Temperature Physics}}
  \textbf{\bibinfo{volume}{209}}, \bibinfo{pages}{337--345}
  (\bibinfo{year}{2022}).

\bibitem{2023ITAS...3359334B}
\bibinfo{author}{{Boyd}, S.~T.~P.}, \bibinfo{author}{{Hines}, N.~R.},
  \bibinfo{author}{{Kavner}, A.~R.~L.} \& \bibinfo{author}{{Kim}, G.-B.}
\newblock \bibinfo{title}{{Development of Fast Decay-Energy Spectroscopy With
  Magnetic Microcalorimeters}}.
\newblock \emph{\bibinfo{journal}{IEEE Transactions on Applied
  Superconductivity}} \textbf{\bibinfo{volume}{33}}, \bibinfo{pages}{3259334}
  (\bibinfo{year}{2023}).

\bibitem{2024ApPhL.124c2601K}
\bibinfo{author}{{Krantz}, M.} \emph{et~al.}
\newblock \bibinfo{title}{{Magnetic microcalorimeter with paramagnetic
  temperature sensors and integrated dc-SQUID readout for high-resolution x-ray
  emission spectroscopy}}.
\newblock \emph{\bibinfo{journal}{Applied Physics Letters}}
  \textbf{\bibinfo{volume}{124}}, \bibinfo{pages}{032601}
  (\bibinfo{year}{2024}).
\newblock \eprint{2310.08698}.

\bibitem{Note1}
\bibinfo{note}{And, thus, minimizing complications involved in fabricating and
  manipulating the wires and SQUIDs necessary for the desired sizes and numbers
  of absorbers. Additionally, SQUIDs may also only tolerate magnetic fields of
  certain strengths, and are also more sensitive to unintended inductances and
  currents.}

\bibitem{2014ARPC...65...83S}
\bibinfo{author}{{Schirhagl}, R.}, \bibinfo{author}{{Chang}, K.},
  \bibinfo{author}{{Loretz}, M.} \& \bibinfo{author}{{Degen}, C.~L.}
\newblock \bibinfo{title}{{Nitrogen-Vacancy Centers in Diamond: Nanoscale
  Sensors for Physics and Biology}}.
\newblock \emph{\bibinfo{journal}{Annual Review of Physical Chemistry}}
  \textbf{\bibinfo{volume}{65}}, \bibinfo{pages}{83--105}
  (\bibinfo{year}{2014}).

\bibitem{doherty2013nitrogen}
\bibinfo{author}{Doherty, M.~W.} \emph{et~al.}
\newblock \bibinfo{title}{The nitrogen-vacancy colour centre in diamond}.
\newblock \emph{\bibinfo{journal}{Physics Reports}}
  \textbf{\bibinfo{volume}{528}}, \bibinfo{pages}{1--45}
  (\bibinfo{year}{2013}).

\bibitem{2019Nanop...8..209L}
\bibinfo{author}{{Levine}, E.~V.} \emph{et~al.}
\newblock \bibinfo{title}{{Principles and techniques of the quantum diamond
  microscope}}.
\newblock \emph{\bibinfo{journal}{Nanophotonics}} \textbf{\bibinfo{volume}{8}},
  \bibinfo{pages}{209} (\bibinfo{year}{2019}).
\newblock \eprint{1910.00061}.

\bibitem{2013Natur.500...54K}
\bibinfo{author}{{Kucsko}, G.} \emph{et~al.}
\newblock \bibinfo{title}{{Nanometre-scale thermometry in a living cell}}.
\newblock \emph{\bibinfo{journal}{\nat}} \textbf{\bibinfo{volume}{500}},
  \bibinfo{pages}{54--58} (\bibinfo{year}{2013}).
\newblock \eprint{1304.1068}.

\bibitem{2017NanoL..17.1496B}
\bibinfo{author}{{Barson}, M. S.~J.} \emph{et~al.}
\newblock \bibinfo{title}{{Nanomechanical Sensing Using Spins in Diamond}}.
\newblock \emph{\bibinfo{journal}{Nano Letters}} \textbf{\bibinfo{volume}{17}},
  \bibinfo{pages}{1496--1503} (\bibinfo{year}{2017}).
\newblock \eprint{1612.05325}.

\bibitem{2017Natur.549..252G}
\bibinfo{author}{{Gross}, I.} \emph{et~al.}
\newblock \bibinfo{title}{{Real-space imaging of non-collinear
  antiferromagnetic order with a single-spin magnetometer}}.
\newblock \emph{\bibinfo{journal}{\nat}} \textbf{\bibinfo{volume}{549}},
  \bibinfo{pages}{252--256} (\bibinfo{year}{2017}).
\newblock \eprint{2011.12399}.

\bibitem{2018PhRvL.121x6402M}
\bibinfo{author}{{Mittiga}, T.} \emph{et~al.}
\newblock \bibinfo{title}{{Imaging the Local Charge Environment of
  Nitrogen-Vacancy Centers in Diamond}}.
\newblock \emph{\bibinfo{journal}{\prl}} \textbf{\bibinfo{volume}{121}},
  \bibinfo{pages}{246402} (\bibinfo{year}{2018}).
\newblock \eprint{1809.01668}.

\bibitem{2019ApPhL.114w1103W}
\bibinfo{author}{{Webb}, J.~L.} \emph{et~al.}
\newblock \bibinfo{title}{{Nanotesla sensitivity magnetic field sensing using a
  compact diamond nitrogen-vacancy magnetometer}}.
\newblock \emph{\bibinfo{journal}{Applied Physics Letters}}
  \textbf{\bibinfo{volume}{114}}, \bibinfo{pages}{231103}
  (\bibinfo{year}{2019}).

\bibitem{hsieh2019imaging}
\bibinfo{author}{Hsieh, S.} \emph{et~al.}
\newblock \bibinfo{title}{Imaging stress and magnetism at high pressures using
  a nanoscale quantum sensor}.
\newblock \emph{\bibinfo{journal}{Science}} \textbf{\bibinfo{volume}{366}},
  \bibinfo{pages}{1349--1354} (\bibinfo{year}{2019}).

\bibitem{block2021optically}
\bibinfo{author}{Block, M.} \emph{et~al.}
\newblock \bibinfo{title}{Optically enhanced electric field sensing using
  nitrogen-vacancy ensembles}.
\newblock \emph{\bibinfo{journal}{Physical Review Applied}}
  \textbf{\bibinfo{volume}{16}}, \bibinfo{pages}{024024}
  (\bibinfo{year}{2021}).

\bibitem{su2021search}
\bibinfo{author}{Su, H.} \emph{et~al.}
\newblock \bibinfo{title}{Search for exotic spin-dependent interactions with a
  spin-based amplifier}.
\newblock \emph{\bibinfo{journal}{Science Advances}}
  \textbf{\bibinfo{volume}{7}}, \bibinfo{pages}{eabi9535}
  (\bibinfo{year}{2021}).

\bibitem{2024Natur.627...73B}
\bibinfo{author}{{Bhattacharyya}, P.} \emph{et~al.}
\newblock \bibinfo{title}{{Imaging the Meissner effect in hydride
  superconductors using quantum sensors}}.
\newblock \emph{\bibinfo{journal}{\nat}} \textbf{\bibinfo{volume}{627}},
  \bibinfo{pages}{73--79} (\bibinfo{year}{2024}).
\newblock \eprint{2306.03122}.

\bibitem{jiao2021experimental}
\bibinfo{author}{Jiao, M.}, \bibinfo{author}{Guo, M.}, \bibinfo{author}{Rong,
  X.}, \bibinfo{author}{Cai, Y.-F.} \& \bibinfo{author}{Du, J.}
\newblock \bibinfo{title}{Experimental constraint on an exotic parity-odd
  spin-and velocity-dependent interaction with a single electron spin quantum
  sensor}.
\newblock \emph{\bibinfo{journal}{Physical Review Letters}}
  \textbf{\bibinfo{volume}{127}}, \bibinfo{pages}{010501}
  (\bibinfo{year}{2021}).

\bibitem{glenn2017micrometer}
\bibinfo{author}{Glenn, D.~R.} \emph{et~al.}
\newblock \bibinfo{title}{Micrometer-scale magnetic imaging of geological
  samples using a quantum diamond microscope}.
\newblock \emph{\bibinfo{journal}{Geochemistry, Geophysics, Geosystems}}
  \textbf{\bibinfo{volume}{18}}, \bibinfo{pages}{3254--3267}
  (\bibinfo{year}{2017}).

\bibitem{thiel2019probing}
\bibinfo{author}{Thiel, L.} \emph{et~al.}
\newblock \bibinfo{title}{Probing magnetism in 2d materials at the nanoscale
  with single-spin microscopy}.
\newblock \emph{\bibinfo{journal}{Science}} \textbf{\bibinfo{volume}{364}},
  \bibinfo{pages}{973--976} (\bibinfo{year}{2019}).

\bibitem{ariyaratne2018nanoscale}
\bibinfo{author}{Ariyaratne, A.}, \bibinfo{author}{Bluvstein, D.},
  \bibinfo{author}{Myers, B.~A.} \& \bibinfo{author}{Jayich, A. C.~B.}
\newblock \bibinfo{title}{Nanoscale electrical conductivity imaging using a
  nitrogen-vacancy center in diamond}.
\newblock \emph{\bibinfo{journal}{Nature communications}}
  \textbf{\bibinfo{volume}{9}}, \bibinfo{pages}{2406} (\bibinfo{year}{2018}).

\bibitem{ajoy2015atomic}
\bibinfo{author}{Ajoy, A.}, \bibinfo{author}{Bissbort, U.},
  \bibinfo{author}{Lukin, M.~D.}, \bibinfo{author}{Walsworth, R.~L.} \&
  \bibinfo{author}{Cappellaro, P.}
\newblock \bibinfo{title}{Atomic-scale nuclear spin imaging using
  quantum-assisted sensors in diamond}.
\newblock \emph{\bibinfo{journal}{Physical Review X}}
  \textbf{\bibinfo{volume}{5}}, \bibinfo{pages}{011001} (\bibinfo{year}{2015}).

\bibitem{rajendran2017method}
\bibinfo{author}{Rajendran, S.}, \bibinfo{author}{Zobrist, N.},
  \bibinfo{author}{Sushkov, A.~O.}, \bibinfo{author}{Walsworth, R.} \&
  \bibinfo{author}{Lukin, M.}
\newblock \bibinfo{title}{A method for directional detection of dark matter
  using spectroscopy of crystal defects}.
\newblock \emph{\bibinfo{journal}{Physical Review D}}
  \textbf{\bibinfo{volume}{96}}, \bibinfo{pages}{035009}
  (\bibinfo{year}{2017}).

\bibitem{chigusa2023light}
\bibinfo{author}{Chigusa, S.}, \bibinfo{author}{Hazumi, M.},
  \bibinfo{author}{Herbschleb, E.~D.}, \bibinfo{author}{Mizuochi, N.} \&
  \bibinfo{author}{Nakayama, K.}
\newblock \bibinfo{title}{Light dark matter search with nitrogen-vacancy
  centers in diamonds}.
\newblock \emph{\bibinfo{journal}{arXiv preprint arXiv:2302.12756}}
  (\bibinfo{year}{2023}).

\bibitem{kashem2025multiplexed}
\bibinfo{author}{Kashem, M. S.~B.} \emph{et~al.}
\newblock \bibinfo{title}{Multiplexed quantum sensing reveals coordinated
  thermomagnetic regulation of mitochondria}.
\newblock \emph{\bibinfo{journal}{bioRxiv}} \bibinfo{pages}{2025--07}
  (\bibinfo{year}{2025}).

\bibitem{NIST}
\bibinfo{author}{Hubbell, J.},  \& \bibinfo{author}{Seltzer, S.~M.}
\newblock \bibinfo{title}{{Tables of X-Ray Mass Attenuation Coefficients and
  Mass Energy-Absorption Coefficients from 1 keV to 20 MeV for Elements Z = 1
  to 92 and 48 Additional Substances of Dosimetric Interest}}
  (\bibinfo{year}{2004}).

\bibitem{2007mmlt.book.....P}
\bibinfo{author}{{Pobell}, F.}
\newblock \emph{\bibinfo{title}{{Matter and Methods at Low Temperatures}}}
  (\bibinfo{year}{2007}).

\bibitem{1962PhRv..128.1622E}
\bibinfo{author}{{Ehrenreich}, H.} \& \bibinfo{author}{{Philipp}, H.~R.}
\newblock \bibinfo{title}{{Optical Properties of Ag and Cu}}.
\newblock \emph{\bibinfo{journal}{Physical Review}}
  \textbf{\bibinfo{volume}{128}}, \bibinfo{pages}{1622--1629}
  (\bibinfo{year}{1962}).

\bibitem{1961PhRv..121.1100T}
\bibinfo{author}{{Taft}, E.~A.} \& \bibinfo{author}{{Philipp}, H.~R.}
\newblock \bibinfo{title}{{Optical Constants of Silver}}.
\newblock \emph{\bibinfo{journal}{Physical Review}}
  \textbf{\bibinfo{volume}{121}}, \bibinfo{pages}{1100--1103}
  (\bibinfo{year}{1961}).

\bibitem{smeltzer2009robust}
\bibinfo{author}{Smeltzer, B.}, \bibinfo{author}{McIntyre, J.} \&
  \bibinfo{author}{Childress, L.}
\newblock \bibinfo{title}{Robust control of individual nuclear spins in
  diamond}.
\newblock \emph{\bibinfo{journal}{Physical Review A}}
  \textbf{\bibinfo{volume}{80}}, \bibinfo{pages}{050302}
  (\bibinfo{year}{2009}).

\bibitem{Note2}
\bibinfo{note}{The contrast $C$ refers to the percent change in fluorescence
  between the ambient value of fluorescence and the decreased fluorescence at
  the bottom of the Lorentzian dip of the resonance. For example, the left
  peaks in Group I of Fig.~\ref {fig:fig2}c show contrasts of a little less
  than 2.5~\%.}

\bibitem{dreau2011avoiding}
\bibinfo{author}{Dr{\'e}au, A.} \emph{et~al.}
\newblock \bibinfo{title}{Avoiding power broadening in optically detected
  magnetic resonance of single nv defects for enhanced dc magnetic field
  sensitivity}.
\newblock \emph{\bibinfo{journal}{Physical Review B}}
  \textbf{\bibinfo{volume}{84}}, \bibinfo{pages}{195204}
  (\bibinfo{year}{2011}).

\bibitem{2020RvMP...92a5004B}
\bibinfo{author}{{Barry}, J.~F.} \emph{et~al.}
\newblock \bibinfo{title}{{Sensitivity optimization for NV-diamond
  magnetometry}}.
\newblock \emph{\bibinfo{journal}{Reviews of Modern Physics}}
  \textbf{\bibinfo{volume}{92}}, \bibinfo{pages}{015004}
  (\bibinfo{year}{2020}).
\newblock \eprint{1903.08176}.

\bibitem{he2023quasi}
\bibinfo{author}{He, G.} \emph{et~al.}
\newblock \bibinfo{title}{Quasi-floquet prethermalization in a disordered
  dipolar spin ensemble in diamond}.
\newblock \emph{\bibinfo{journal}{Physical Review Letters}}
  \textbf{\bibinfo{volume}{131}}, \bibinfo{pages}{130401}
  (\bibinfo{year}{2023}).

\bibitem{2010ApPhL..97x1901H}
\bibinfo{author}{{Hadden}, J.~P.} \emph{et~al.}
\newblock \bibinfo{title}{{Strongly enhanced photon collection from diamond
  defect centers under microfabricated integrated solid immersion lenses}}.
\newblock \emph{\bibinfo{journal}{Applied Physics Letters}}
  \textbf{\bibinfo{volume}{97}}, \bibinfo{pages}{241901}
  (\bibinfo{year}{2010}).
\newblock \eprint{1006.2093}.

\bibitem{Note3}
\bibinfo{note}{The second term in equation \ref {eq:key}, dT/dB, also improves
  with smaller absorber size (on average, improving with the square root of
  $L$) until about 2 or 3~$\mu $m, after which it starts worsening with
  decreasing absorber size. However, we only see this slowly take effect,
  leading to the turnover being at around 1~$\mu $m, as it needs to override
  the main trend previously described.}

\bibitem{2020ApJ...891...70A}
\bibinfo{author}{{Abarr}, Q.} \emph{et~al.}
\newblock \bibinfo{title}{{Observations of a GX 301-2 Apastron Flare with the
  X-Calibur Hard X-Ray Polarimeter Supported by NICER, the Swift XRT and BAT,
  and Fermi GBM}}.
\newblock \emph{\bibinfo{journal}{\apj}} \textbf{\bibinfo{volume}{891}},
  \bibinfo{pages}{70} (\bibinfo{year}{2020}).
\newblock \eprint{2001.03581}.

\bibitem{2022APh...14302749A}
\bibinfo{author}{{Abarr}, Q.} \emph{et~al.}
\newblock \bibinfo{title}{{Performance of the X-Calibur hard X-ray polarimetry
  mission during its 2018/19 long-duration balloon flight}}.
\newblock \emph{\bibinfo{journal}{Astroparticle Physics}}
  \textbf{\bibinfo{volume}{143}}, \bibinfo{pages}{102749}
  (\bibinfo{year}{2022}).
\newblock \eprint{2204.09761}.

\bibitem{Note4}
\bibinfo{note}{Gau et al., in progress}.

\bibitem{2020apra.prop...44K}
\bibinfo{author}{{Krawczynski}, H.}
\newblock \bibinfo{title}{{Test of a Novel Mini-Dilution Refrigerator and a
  SLEDGEHAMMER X-ray Microcalorimeter Array on a Conventional Balloon Flight}}.
\newblock \bibinfo{howpublished}{NASA Proposal id. 20-APRA20-44}
  (\bibinfo{year}{2020}).

\bibitem{2023JATIS...9b4006S}
\bibinfo{author}{{Shirazi}, F.} \emph{et~al.}
\newblock \bibinfo{title}{{511-CAM mission: a pointed 511 keV gamma-ray
  telescope with a focal plane detector made of stacked transition edge sensor
  microcalorimeter arrays}}.
\newblock \emph{\bibinfo{journal}{Journal of Astronomical Telescopes,
  Instruments, and Systems}} \textbf{\bibinfo{volume}{9}},
  \bibinfo{pages}{024006} (\bibinfo{year}{2023}).
\newblock \eprint{2206.14652}.

\bibitem{2018PhRvX...8a1042W}
\bibinfo{author}{{Wang}, N.} \emph{et~al.}
\newblock \bibinfo{title}{{Magnetic Criticality Enhanced Hybrid Nanodiamond
  Thermometer under Ambient Conditions}}.
\newblock \emph{\bibinfo{journal}{Physical Review X}}
  \textbf{\bibinfo{volume}{8}}, \bibinfo{pages}{011042} (\bibinfo{year}{2018}).
\newblock \eprint{1707.02885}.

\bibitem{2018JLTP..193..435B}
\bibinfo{author}{{Boyd}, S.~T.~P.} \emph{et~al.}
\newblock \bibinfo{title}{{Integrated SQUID/Sensor Metallic Magnetic
  Microcalorimeter for Gamma-Ray Spectroscopy}}.
\newblock \emph{\bibinfo{journal}{Journal of Low Temperature Physics}}
  \textbf{\bibinfo{volume}{193}}, \bibinfo{pages}{435--440}
  (\bibinfo{year}{2018}).

\bibitem{2020PhRvL.125n2503S}
\bibinfo{author}{{Sikorsky}, T.} \emph{et~al.}
\newblock \bibinfo{title}{{Measurement of the $^{229}$Th Isomer Energy with a
  Magnetic Microcalorimeter}}.
\newblock \emph{\bibinfo{journal}{\prl}} \textbf{\bibinfo{volume}{125}},
  \bibinfo{pages}{142503} (\bibinfo{year}{2020}).
\newblock \eprint{2005.13340}.

\bibitem{2020JLTP..199.1055K}
\bibinfo{author}{{Kim}, G.~B.} \emph{et~al.}
\newblock \bibinfo{title}{{A New Measurement of the 60 keV Emission from Am-241
  Using Metallic Magnetic Calorimeters}}.
\newblock \emph{\bibinfo{journal}{Journal of Low Temperature Physics}}
  \textbf{\bibinfo{volume}{199}}, \bibinfo{pages}{1055--1061}
  (\bibinfo{year}{2020}).

\bibitem{2022JLTP..209.1119H}
\bibinfo{author}{{Herbst}, M.} \emph{et~al.}
\newblock \bibinfo{title}{{Numerical Calculation of the Thermodynamic
  Properties of Silver Erbium Alloys for Use in Metallic Magnetic
  Calorimeters}}.
\newblock \emph{\bibinfo{journal}{Journal of Low Temperature Physics}}
  \textbf{\bibinfo{volume}{209}}, \bibinfo{pages}{1119--1127}
  (\bibinfo{year}{2022}).

\bibitem{2013ITAS...23Q0605S}
\bibinfo{author}{{Stevenson}, T.~R.} \emph{et~al.}
\newblock \bibinfo{title}{{Superconducting Effects in Optimization of Magnetic
  Penetration Thermometers for X-Ray Microcalorimeters}}.
\newblock \emph{\bibinfo{journal}{IEEE Transactions on Applied
  Superconductivity}} \textbf{\bibinfo{volume}{23}},
  \bibinfo{pages}{2300605--2300605} (\bibinfo{year}{2013}).

\bibitem{2020arXiv201012076A}
\bibinfo{author}{{Abeln}, A.} \emph{et~al.}
\newblock \bibinfo{title}{{Conceptual Design of BabyIAXO, the intermediate
  stage towards the International Axion Observatory}}.
\newblock \emph{\bibinfo{journal}{arXiv e-prints}}
  \bibinfo{pages}{arXiv:2010.12076} (\bibinfo{year}{2020}).
\newblock \eprint{2010.12076}.

\bibitem{levine2019principles}
\bibinfo{author}{Levine, E.~V.} \emph{et~al.}
\newblock \bibinfo{title}{Principles and techniques of the quantum diamond
  microscope}.
\newblock \emph{\bibinfo{journal}{Nanophotonics}} \textbf{\bibinfo{volume}{8}},
  \bibinfo{pages}{1945--1973} (\bibinfo{year}{2019}).

\bibitem{PhysRevB.102.134210}
\bibinfo{author}{Bauch, E.} \emph{et~al.}
\newblock \bibinfo{title}{Decoherence of ensembles of nitrogen-vacancy centers
  in diamond}.
\newblock \emph{\bibinfo{journal}{Phys. Rev. B}}
  \textbf{\bibinfo{volume}{102}}, \bibinfo{pages}{134210}
  (\bibinfo{year}{2020}).
\newblock \urlprefix\url{https://link.aps.org/doi/10.1103/PhysRevB.102.134210}.

\bibitem{gruber1997scanning}
\bibinfo{author}{Gruber, A.} \emph{et~al.}
\newblock \bibinfo{title}{Scanning confocal optical microscopy and magnetic
  resonance on single defect centers}.
\newblock \emph{\bibinfo{journal}{Science}} \textbf{\bibinfo{volume}{276}},
  \bibinfo{pages}{2012--2014} (\bibinfo{year}{1997}).

\bibitem{Note5}
\bibinfo{note}{Https://www.axiomoptics.com/products/kinetix/}.

\bibitem{2019MS&E..502a2134C}
\bibinfo{author}{{Chase}, S.~T.}, \bibinfo{author}{{Brien}, T.~L.~R.},
  \bibinfo{author}{{Doyle}, S.~M.} \& \bibinfo{author}{{Kenny}, L.~C.}
\newblock \bibinfo{title}{{Pre-cooling a $^{3}$He/$^{4}$He dilutor module with
  a sealed closed-cycle continuous cooler}}.
\newblock In \emph{\bibinfo{booktitle}{Materials Science and Engineering
  Conference Series}}, vol. \bibinfo{volume}{502} of
  \emph{\bibinfo{series}{Materials Science and Engineering Conference Series}},
  \bibinfo{pages}{012134} (\bibinfo{year}{2019}).

\bibitem{2023JATIS...9a4002K}
\bibinfo{author}{{Kislat}, F.} \emph{et~al.}
\newblock \bibinfo{title}{{ASCENT: a balloon-borne hard x-ray imaging
  spectroscopy telescope using transition edge sensor microcalorimeter
  detectors}}.
\newblock \emph{\bibinfo{journal}{Journal of Astronomical Telescopes,
  Instruments, and Systems}} \textbf{\bibinfo{volume}{9}},
  \bibinfo{pages}{014002} (\bibinfo{year}{2023}).
\newblock \eprint{2301.01525}.

\end{thebibliography}

%%%%%%%%%% Merge with supplemental materials %%%%%%%%%%
\widetext
\clearpage
\begin{center}
\textbf{\large Supplementary Materials}
\end{center}
%%%%%%%%%% Merge with supplemental materials %%%%%%%%%%
%%%%%%%%%% Prefix a "S" to all equations, figures, tables and reset the counter %%%%%%%%%%
\setcounter{equation}{0}
\setcounter{figure}{0}
\setcounter{table}{0}
\setcounter{page}{1}
\makeatletter
\renewcommand{\theequation}{S\arabic{equation}}
\renewcommand{\thefigure}{S\arabic{figure}}
% \renewcommand{\bibnumfmt}[1]{[S#1]}
% \renewcommand{\citenumfont}[1]{S#1}
%%%%%%%%%% Prefix a "S" to all equations, figures, tables and reset the counter %%%%%%%%%%

\maketitle

\section{Further details of the experimental setup}
\label{s:setupagain}

As described in the main text, the distance between the absorbers and the diamond is $\sim$100~nm. 

A laser is required to optically address the NV centers. 
To achieve optimal sensitivity, as calculated in Figure~3b, a laser intensity of $\sim$50~W/mm$^2$ will be used. 
However, as described above, the laser can be aimed at a sufficiently high angle of incidence such that it undergoes total internal reflection at the top surface of the diamond, as schematically depicted in Figure~1b.
In order to achieve total internal reflection at the interface of the diamond with the silver, then, the incoming 532~nm (green) laser would only need to approach that surface at an angle from the normal to the interface of no less than a couple degrees, given the very low refractive index of silver ($n\sim0.1$) \cite{1961PhRv..121.1100T} compared to that of diamond ($n\sim2.4$). 

Then, using the same experimental setup as used by a subset of the current authors, the infinity-corrected objective lens (OL) can be positioned $\sim$15~mm below the bottom of the diamond, with $\sim$5~mm of that distance being between the diamond and the cryostat window and the remaining distance being between the cryostat window and the objective lens outside. 

Finally, the distance between the OL and the CCD, with the filter and the tube lens inserted in between the two, would be about $\sim$200~mm.

%%%%%%%%%%%%%%%%%%%%%%%%%%%%%%%%%%%%%%%%%%%%%%%%%%%%%%%%%%%%%%%%%%%%%%%%%%%%%%%%%

\section{Calculation of NV volume sensitivity}
\label{s:sense}

The magnetic field volume sensitivity in NV ESR measurements, as derived in~\cite{levine2019principles}, can be expressed as:
\begin{align}
    \eta \approx \frac{8\pi}{3\sqrt{3}} \frac{1}{\gamma} \frac{\delta\nu}{C \sqrt{R}} ,  \label{eq:sensitivity_Sup}
\end{align}
which is identical to Equation~3 in the main text.
In a typical NV ESR measurement, the resonance signal follows a Lorentzian profile~\cite{dreau2011avoiding}, characterized by the linewidth $\delta\nu$, contrast $C$, and photon count rate $R$, which are given by the equations below.

\subsection{Linewidth}

The linewidth, $\delta\nu$, can be expressed as
\begin{align}
    \delta\nu = \frac{1}{2\pi} \sqrt{(\Gamma^*_2 + \Gamma_c )^2 + \Omega^2 \frac{\Gamma^*_2 + \Gamma_c}{2\Gamma_1 + \Gamma_p}},
    \label{eq:linewidth}
\end{align}
where $\Omega$, as in Figure 3, is the microwave Rabi frequency (the power).
In addition to the NV spin-lattice intrinsic relaxation rate $\Gamma_1$ and dephasing rate $\Gamma^*_2$, the optical pumping phenomenologically induces a polarization rate $\Gamma_p$ and a decoherence rate $\Gamma_c$ that act in conjunction with $\Gamma_1$ and $\Gamma^*_2$, respectively.
Typically, the intrinsic dephasing rate $\Gamma^*_2$ is proportion to the nitrogen concentration in diamond, with a proportionality factor of $0.1~\mathrm{\mu s}^{-1} / \mathrm{ppm}$~\cite{PhysRevB.102.134210}.
With [N] = 20 $\mathrm{ppm}$ from Table~1 of the main text, we can thus take $\Gamma^*_2 = 2~{\rm \mu s^{-1}}$.
The relaxation rate $\Gamma_1 \approx 1~\mathrm{ms}^{-1}$~\cite{dreau2011avoiding} is significantly slower than other decay rates.

The optical induced rates exhibit a standard saturation behavior with respect to the laser power.
By defining $s = I_\mathrm{exp}/I_\mathrm{sat}$ as the ratio of the experimental laser intensity $I_\mathrm{exp}$ to the saturation intensity $I_\mathrm{sat}$, the polarization rate $\Gamma_p$ and decoherence rate $\Gamma_c$ can be written as
\begin{align}
    \Gamma_p = \Gamma^{\infty}_p \frac{s}{1+s} , \,\,\,\,\,\,\,\,\,\,
    \Gamma_c = \Gamma^{\infty}_c \frac{s}{1+s} ,
    \label{eq:rate}
\end{align}
where $\Gamma^{\infty}_p \approx 5~\mathrm{\mu s}^{-1}$ and $\Gamma^{\infty}_c \approx 80~\mathrm{\mu s}^{-1}$~\cite{dreau2011avoiding} are the rates at laser saturation.
The saturation intensity $I_\mathrm{sat}$ for NV centers is approximately 1~--~3~$\mathrm{mW/\mu m^2}$~\cite{2020RvMP...92a5004B, gruber1997scanning}.
We take $I_\mathrm{sat} = 2~\mathrm{mW/\mu m^2} = 2000~\mathrm{W/mm^2}$ in our calculation.

\subsection{Contrast}

The contrast $C$ is given by
\begin{align}
    C = \Theta \frac{\Omega^2}{\Omega^2 + (2\Gamma_1 + \Gamma_p) (\Gamma^*_2 + \Gamma_c)} ,
    \label{eq:contrast}
\end{align}
with $\Theta =~$Contrast$_{\mathrm{max}}/2 = 3.75\%$ serving as the overall normalization factor.

\subsection{Photon count rate}

The photon count rate $R$ also follows a standard saturation behavior,
\begin{align}
    R = R^{\infty} \frac{s}{1+s} ,
    \label{eq:counts}
\end{align}
where the saturation count rate is $R^{\infty} = 1.76 \times 10^5 \frac{\mathrm{\mu m}^{-3}}{ppm} \cdot [NV] \cdot R_\mathrm{sat} = 7.92 \times 10^{10}~\mathrm{\mu m}^{-3}~\mathrm{s}^{-1}$, calculated with the diamond parameters provided in Table~I in the main text.

Using the three ESR parameters calculated as above, the NV magnetic field volume sensitivity is given by Equation~\ref{eq:sensitivity_Sup}.
Figure~3b of the main text shows how the volume sensitivity varies with the (experimental) laser intensity and microwave strength.
An optimal minimum sensitivity value, $\eta_\mathrm{opt} \approx 27.25~{\rm nT}~{\rm Hz}^{-\frac{1}{2}}~\mathrm{\mu m}^{\frac{3}{2}}$, is achieved at a laser intensity of $I \approx 51.2~{\rm W}\cdot{\rm mm}^{-2}$ and a microwave strength of $\Omega \approx (2\pi) \times 0.16$~MHz.

\section{Calculation of optimal NV center volume}
\label{s:vol}

As indicated by Equation~2 in the main text, the magnetic field from the absorber pad will decrease cubically with increasing distance between the absorbers and the NV centers in diamond.
Additionally, the length of the absorber pad $L$ plays a crucial role in this relationship.
Therefore, in order to calculate the optimal sensitivity of the NV centers, it is important to carefully consider the widths and thickness of the NV layers that would meaningfully contribute to measuring the magnetic field changes due to the temperature transient.

\begin{figure*}[h]
\begin{center}
    \includegraphics[width=.9\linewidth]{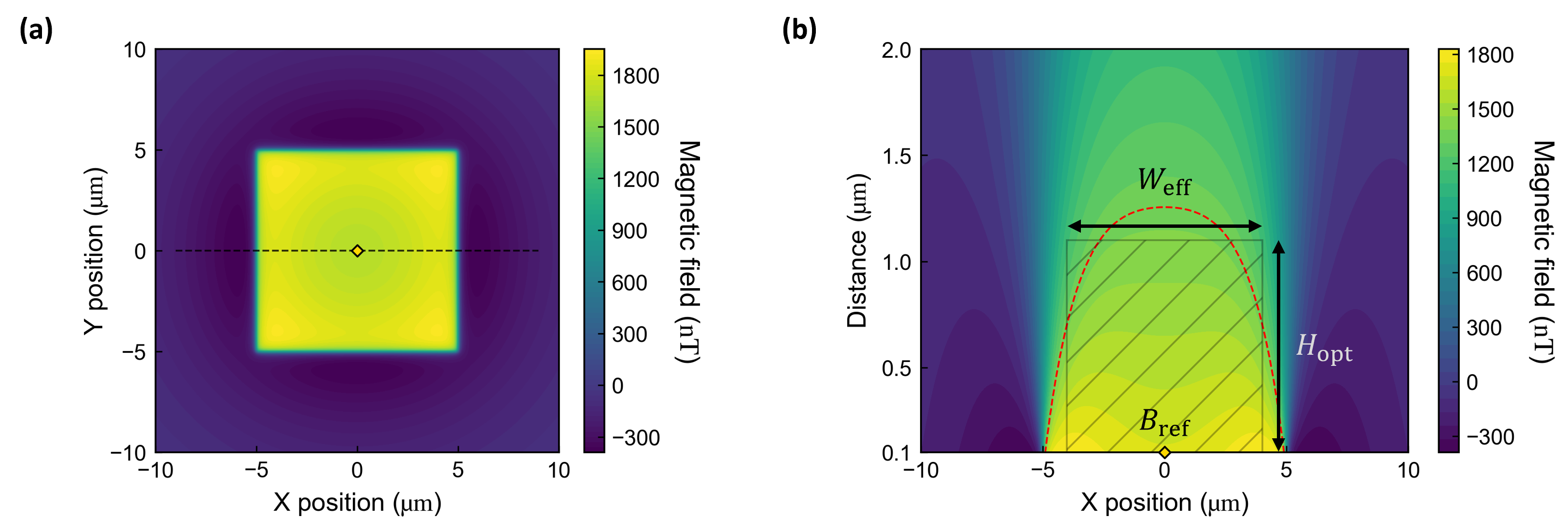}
\end{center}
\vspace*{-0.3cm}
\caption{{\bf Principle of NV layer volume selection.}
{\bf (a)} The calculated distribution of magnetic field change at the location of the NV centers in diamond, 0.1~$\mu$m beneath a $10~\mu \rm m \times 10~\mu \rm m \times 5~\mu \rm m$ (for example) absorber pad, given a $1$~mK increase in the temperature of the absorber.
Calculations, such as the one done for this figure, were done for all of the absorber widths shown in Figure 4.
{\bf (b)} Magnetic field strength at the depth of a given NV layer as a function of the X-coordinate and of the distance between the layer and the absorber pad, taken along the dashed line-cut axis shown in (a).
The reference field magnitude, $B_\mathrm{ref}$ (measured at the location given by the diamond), is defined as the field strength at a position of X = 0 at a distance of 0.1~$\mu$m below the absorber.
The red dashed line marks the region where the field magnitude stays above 80\% of $B_\mathrm{ref}$.
The hatched area, with width $W_{\mathrm{eff}}$ and height $H_{\mathrm{opt}}$, is then taken to be the effective volume within which we accumulate NV signals for each magnetic field measurement.
}
 \label{fig:figS1}
\end{figure*}

We consider an absorber pad with dimensions $10~\mu \rm m \times 10~\mu \rm m \times 5~\mu \rm m$ here as an example.
At the closest distance of the diamond (0.1~$\mu$m) to the absorber pad, where the distance is much smaller than the length of the pad, the variation of magnetic field within the diamond sample is negligible, as seen in Fig.~\ref{fig:figS1}(a).
We take the magnitude at the center, $B_\mathrm{ref}$ (Fig.~\ref{fig:figS1}(b)), as the reference field strength.
Then, along a line through the the center of the absorber pad (the black dashed line in Fig.~\ref{fig:figS1}(a)), we simulate the magnetic field in the diamond at different distances from the absorber pad: this is plotted in Fig.~\ref{fig:figS1}(b).
By taking field strengths greater than 80\% of $B_\mathrm{ref}$ as acceptable strengths, we draw an 80\% line (the red dashed line in Fig.~\ref{fig:figS1}(b)).
The effective NV layer sensing height, $H_{opt}$, is first determined by the distance at which the width across the region given by the red line does not decrease below half the pad length. 
Then, the effective NV layer sensing width, $W_\mathrm{eff}$, is set by the average value of all the widths (across the region given by the dashed red line) within $H_\mathrm{opt}$. 
Consequently, the volume in the NV layer of the diamond that will be counted towards the total sensitivity below a given absorber pad is calculated with $V_{\mathrm{opt}} =  W_{\mathrm{eff}} \times W_{\mathrm{eff}} \times H_{\mathrm{opt}}$.

Finally, the effective magnetic field strength due to the pad and present throughout this volume is determined by averaging the field: 
\begin{align}
    B_{\mathrm{eff}} =  \frac{1}{V_{\mathrm{opt}}} \iiint_{V_{\mathrm{opt}}} B(\mathbf{r})~\mathrm{d}V \approx 0.9 * B_\mathrm{ref},
    \label{eq:volume}
\end{align}
with $B_{\mathrm{eff}}$ typically around 90\% of $B_\mathrm{ref}$.

%%%%%%%%%%%%%%%%%%%%%%%%%%%%%%%%%%%%%%%%%%%%%%%%%%%%%%%%%%%%%%%%%%%%%%%%%%%%%%%%%

\section{Other considerations/factors for the energy resolution}

Given the characteristic absorber size used above, the energy resolution due purely to the thermal fluctuations (between the thermal subsystems of the spins and the electrons in the paramagnetic absorber), in the absence of other processes, is $\sim$0.095~eV, assuming that Equation 8 from \cite{2003RScI...74.3947F} holds 
(along with other assumptions, such as those from page 196 of \cite{2005cpd..book..151F}).
Even when factoring in the relative effect of interacting magnetic moments and of magnetic Johnson noise as calculated by \cite{2005cpd..book..151F}, the energy resolution due to just these three factors together would still only be $\sim$0.110~eV: altogether, still much smaller than the main effects described in the previous subsections. 

%%%%%%%%%%%%%%%%%%%%%%%%%%%%%%%%%%%%%%%%%%%%%%%%%%%%%%%%%%%%%%%%%%%%%%%%%%%%%%%%%

\section{Time resolution}
\label{s:time}

To estimate the overall time resolution of the NV-MMC detector, we examine the duration of each step in the measurement process following the X-ray incidence onto the absorber.

After the incoming X-ray photoelectrically ejects an electron, that energy is passed back and forth between electrons and phonons, before finally ending up with the electrons and then being diffused, to thermalize the absorber on the timescale of less than 100~ns (see \citep{2000JLTP..121..137E, 2005cpd..book..151F}; the latter source cites $\sim$100~ns for a larger absorber than the ones proposed here).
Then, the paramagnetic spins would receive the energy from the thermalized electrons on the order of 200~ns, according to the Korringa relation provided in \citep{2000JLTP..121..137E, 2005cpd..book..151F} (see also \cite{2018JLTP..193..365K}, which uses the Korringa relation to estimate the overall rise time).

Next, any change in the magnetic field of the absorber, due to its changing temperature as it thermalizes, will modify the ESR spectrum of the NV centers below. 
The timescale of pushing the NV center (with the microwave) between the $|m_s = 0\rangle$ and $|m_s = \pm1\rangle$ states is on the order of hundreds of nanoseconds.
The normal (optically) radiative decay pathway for the NV center from its excited state to the ground state occurs on the order of 10~ns  \cite{dreau2011avoiding}. 
However, if the NV center decays via the non-radiative pathway (as is more likely to happen if the NV center is in the $|m_s = \pm1\rangle$ state), this pathway can take can up to around 200~ns \cite{dreau2011avoiding}.
We would thus expect the overall spectrum to update on a timescale of faster than 1~$\mu$s.

Then, a CCD or CMOS camera (such as one from \footnote{https://www.axiomoptics.com/products/kinetix/}) would permit exposure times down to $\sim$1~$\mu$s and frame rates of $\sim$800~fps ($\sim$1.3~ms).

Finally, the temperature of the pads (through the appropriate thermal coupling to the graphite) would decay with a time constant of $\sim$ 1 to 10~ms back to the ambient temperature of $\sim$35~mK, as described in the main text.
\cite{2005cpd..book..151F} describes using vacuum grease for a longer decay time (10~ms or longer) and epoxy for a shorter decay time (1~ms or shorter).

All in all, we expect the slowest step in the NV-MMC detection process to be that of the absorber returning to its ambient temperature. 
Thus, we predict the overall setup to have a time resolution on the order of milliseconds (either 1~ms or 10~ms), dictated by the decay of the absorber temperature.

%%%%%%%%%%%%%%%%%%%%%%%%%%%%%%%%%%%%%%%%%%%%%%%%%%%%%%%%%%%%%%%%%%%%%%%%%%%%%%%%%

\section{Heating loads}
\label{s:heat}

In this section, we calculate various sources of heat that are generated in the NV-MMC setup.
We discuss ways of designing the setup such that these heat loads do not surpass reasonably-available cooling powers of the millikelvin or the 4~K stages for balloon-borne missions.  

%%%

First, to obtain the ESR spectrum, at most $\sim$10~mW of microwave power would need to be delivered through a stripline positioned at the bottom of the diamond sample. 
In the case of perfect impedance matching and no wire resistance, almost no heat would be generated from the transmission line and dissipated into the diamond. 
However, based on the experience of a subset of the authors, even a conservative estimate would have, at most, $10\%$ of the microwave power ($\sim$1~mW out of $\sim$10~mW) converted to heat at the 4~K stage.

To cool the 4~K stage, a helium-4 tank of volume 28~L has been flown by a subset of the authors for the {\it DR-TES} balloon-borne mission. 
Using a latent heat of evaporation of helium-4 of 2.56~kJ/L \cite{2007mmlt.book.....P}, a heat load of only $\sim$1~mW from the application of the microwave will, by itself, result in the evaporation of the tank only after a much longer time (greater than two years) than the timescale of a long-duration balloon-borne mission (on the scale of days to, at most, months). 
If a larger cooling capacity were needed, one could also consider a larger tank, or other methods of cooling (such as a pulse tube). 
Thus, the cooling power available for the 4~K stage is sufficient for mitigating this possible input of heat.

%%%

Second, the NV centers (reading out an array of size $1~\mathrm{mm}\times1~\mathrm{mm}$) are expected to radiate $\sim$7.9$\times 10^{16}$ red-wavelength photons (given the same width of diamond as described in the main text, populated by the same thickness of NV centers) downwards towards the camera each second. 
The NV centers will thus also emit photons at the same rate upwards towards the absorbers. 
The wavelengths of these red photons range from 630 to 800~nm \citep{2014ARPC...65...83S}, leading to a conservative estimate (using the more energetic end of that range) of a heat load of $\sim$25~mW on the absorber pads due to the NV fluorescence.
To prevent such direct radiation onto the coldest stage, we deposit above the diamond the thin layer ($\sim$50~nm) of silver, having a nearly complete reflectance at the relevant energies (below 2 eV) \cite{1962PhRv..128.1622E, 1961PhRv..121.1100T}.

The thermal conductivity given in \cite{2007mmlt.book.....P} for AGOT graphite is $\sim$2 $\times$ 10$^{-7}$ W/cm/K at $\sim$150~mK. 
Taking the graphite layer width of $\sim$50~nm, the aformentioned interfacial area of $1~\mathrm{mm}\times1~\mathrm{mm}$, and the temperature difference between the 4~K and 35~mK stages, the heat load on the coldest stage would be $\sim$1.6~mW.
Given that various tested cooling options available for balloon flights (a continuous dilution refrigerator, or an adiabatic demagnetization refrigerator) to cool the millikelvin stage to about $\sim$75~mK can sustain a 1~$\mu$W load \citep{2019MS&E..502a2134C, 2023JATIS...9a4002K}, a compromise may have to be made for the array size (a square array of width $\sim$30~$\mu$m would achieve a heat load of this magnitude).
Alternatively, different geometries for the AGOT graphite layer could be explored, such that the interfacial area provides sufficient mechanical and cooling support for the absorbers but does not simply blanket the entire silver layer below---thus reducing the surface area available for heat transfer from the hotter stage. 

%%%%%%%%%%%%%%%%%%%%%%%%%%%%%%%%%%%%%%%%%%%%%%%%%%%%%%%%%%%%%%%%%%%%%%%%%%%%%%%%%

\end{document}